\documentclass[10pt]{iopart}
\usepackage{iopams}
\usepackage{graphicx}
\usepackage{epsfig}
\usepackage{subfigure}
\usepackage{float}

\begin{document}
\title[Cosmological constraints with spectroscopic evolution of cosmic chronometers]{Constraining the expansion rate of the Universe using low-redshift ellipticals as cosmic chronometers}
\author{Michele Moresco$^{1}$, Raul Jimenez$^{2}$, Andrea Cimatti$^{1}$ \& Lucia Pozzetti$^{3}$\\
\vspace*{.1in}
$^1${\it Dipartimento di Astronomia, Universit\'a degli Studi di Bologna, via Ranzani 1, I-40127, Bologna, Italy}\\
$^2${\it ICREA \& Institute of Sciences of the Cosmos (ICC), University of Barcelona, Barcelona 08028, Spain}\\
$^3${\it INAF -- Osservatorio Astronomico di Bologna, via Ranzani 1, I-40127, Bologna, Italy} \\
\vspace*{.1in}
{\rm E-mail: michele.moresco@unibo.it, raul.jimenez@icc.ub.edu, a.cimatti@unibo.it, lucia.pozzetti@oabo.inaf.it}\\}

\begin{abstract}
We present a  new methodology to determine the expansion history of the Universe analyzing the spectral properties of early-type galaxies (ETG), based on the study of the redshift dependence of the 4000 {\AA} break. In this paper we describe the method, explore its robustness using theoretical synthetic stellar population models, and apply it using a SDSS sample of $\sim$14 000 ETGs. Our motivation to look for a new technique has been to minimize the dependence of the cosmic chronometer method on systematic errors. In particular,  as a test of our method, we derive the value of the Hubble constant \mbox{$H_0 = 72.6 \pm 2.9(stat)\pm 2.3(syst)\;\mathrm{km\,Mpc^{-1}s^{-1}}$} (68\% confidence), which is not only fully compatible with the value derived from the Hubble key project, but also with a comparable error budget. Using the SDSS, we also derive, assuming $w=\mathrm{constant}$, a value for the dark energy equation of state parameter \mbox{$w = -1 \pm 0.2(stat)\pm0.3(syst)$}. Given the fact that the SDSS ETG sample only reaches $z \sim 0.3$, this result shows the potential of the method. In future papers we will present results using the high-redshift universe, to yield a determination of $H(z)$ up to $z \sim 1$.
\end{abstract}

\section{Introduction}
\label{sec:intro}
In the last decades many efforts have been done in trying to unveil and understand the expansion history of our Universe. Many different and complementary observations, such as direct supernova measurements of the deceleration parameter (Riess et al. 1998, Perlmutter et al. 1999) as well as indirect measurements based upon a combination of results from the cosmic microwave background (CMB) (Jungman et al. 1996, Miller et al. 1999, De Bernardis et al. 2000, Hanany et al. 2000, Jaffe et al. 2001, Halverson et al. 2002, Mason et al. 2003, Benoit et al. 2003, Goldstein et al. 2003, Spergel et al. 2003, Spergel et al. 2007, Reichardt et al. 2009, Dunkley et al. 2009), large-scale structure (LSS) (Percival et al. 2001, Dodelson et al. 2002), and the Hubble constant (Freedman et al. 2001), pointed out that at present time the expansion of the Universe is accelerating. This could be explained either by considering that gravity at some large scales cannot be described by the standard general relativity and/or that the Universe is filled with some sort of negative-pressure ``dark energy'' that drives the accelerated expansion (Peebles et al. 2003, Padmanabhan 2003, Copeland et al. 2006, Frieman et al. 2008, Linder 2008, Caldwell et al. 2009); either way, it requires new physics beyond general relativity and the standard model of particle physics.

The simplest possibility is to extend Einstein's equation with a cosmological constant, or equivalently, to hypothesize a fluid with an equation-of-state parameter $w\equiv p/\rho = -1$ (with $p$ and $\rho$ the pressure and energy density, respectively).  However, it may well be that the cosmological ``constant'' actually evolves with time, in which case $w\neq -1$, and there are a variety of possibilities to believe that this might be the case (Ratra et al. 1988, Caldwell et al. 1998).  Precise measurement of $w(z)$ (with, in general, a parameterized redshift dependence) or, relatedly, the cosmic expansion history $H(z)$, has thus become a central goal of physical cosmology (Peacock et al. 2006, Albrecht et al. 2006).

At present, the most promising techniques to determine the cosmic expansion history are supernova searches (Riess et al. 1998, Perlmutter et al. 1999), baryon acoustic oscillations (BAO) (Seo et al. 2003, Eisenstein et al. 2005, Percival et al. 2007, Pritchard et al. 2007), weak lensing (Refregier et al. 2003), and galaxy clusters (Haiman et al. 2001).  All these techniques have different strengths and suffer from different systematics and weaknesses.  As argued in the ESO/ESA and Dark Energy Task Force reports (Peacock et al. 2006, Albrecht et al. 2006), robust conclusions about the cosmic expansion history will likely require several independent methods to allow for cross checks and control of systematics.

As suggested by Jimenez \& Loeb (2002), an interesting and complementary approach to this problem is the study of the change in the age of the Universe as a function of redshift. The strength of this method is that it avoids the common weakness of the other techniques, i.e. the reliance on integrated quantities to determine the expansion history. However, to fully exploit this approach, it is required to find a population of standard clocks able to trace the evolution of the relative age of the Universe.\\
Many recent works gave growing observational evidence that the most massive galaxies contain the oldest stellar populations up to redshifts of $z \sim 1-2$ (Dunlop et al. 1996, Spinrad et al. 1997, Cowie et al. 1999, Heavens et al. 2004, Thomas et al. 2005, Cimatti et al. 2008, Thomas et al. 2010) and have less than 1\% of their present stellar mass formed at $z <1$. Studying the mass function of early-type galaxies, it has been found that its massive end evolves very mildly (or not at all) from $z\approx0.7$ to $z=0$ (Cimatti et al. 2006, Pozzetti et al. 2010). This population of massive, red, passive early-type galaxies ({\it ETGs}) consists of the oldest objects in the Universe at each redshift. The differential ages of these galaxies should thus be a good indicator for the rate of change of the age of the Universe as a function of redshift up to $z\sim2$, given that it is commonly accepted that the bulk of their stars was formed at $z>2-3$. There are some works that try to constrain the cosmic expansion history using this approach (Ferreras et al. 2001, Jimenez et al. 2003, Krauss et al. 2003, Capozziello et al. 2004, Stern et al. 2010, Crawford et al. 2010). However, a problem of this technique is that it relies on the determination of a non-observable parameter using spectra and/or spectral energy distributions, i.e. the age of the stellar population, which presents strong and well known degeneracies with other parameters that are evaluated simultaneously, such as the metallicity, the star formation history, the dust content.

The aim of this project is to search for features in the spectra of ETGs that minimize the impact of systematics on the estimate of the relative ages of galaxies, which is the relevant quantity to determine $H(z)$ using the cosmic chronometer method. As we will demonstrate below, the 4000 {\AA} break is a feature correlated almost linearly with age at fixed metallicity. The 4000 {\AA} break (D4000) is a spectral feature produced by the blending of a large number of absorption lines in a narrow wavelength region, that creates a break in the continuum spectrum at 4000 {\AA} restframe. The main contribution to the opacity comes from ionized metals. In hot stars, the metals are multiply ionized and the opacity decreases, so the D4000 will be small for young stellar population and large for old, metal rich galaxies. A measure break amplitude was firstly introduced as the ratio of the average of the flux densities below and above 4000 {\AA} (Hamilton, 1985); he was also the first to suggest that this feature can be used as an age indicator for a galaxy.
Since then, different index definition were proposed, depending on the width of the red and blue regions where the fluxes are evaluated, and many studies have been done on this feature. Balogh et al. (1999) introduced a definition of D4000 with narrower bands, supporting this choice with a lower sensitivity to reddening effects. Bruzual \& Charlot (2003) created a model for computing the spectral evolution of stellar populations (BC03), capable also to trace the dependence of D4000 on age and metallicity as a function of age, reddening and star formation history. Kauffmann et al. (2003) found, from SDSS observations, that the distribution of D4000 is strongly bimodal, showing a clear division between galaxies dominated by old stellar population and galaxies with more recent star formation.

In this paper we analyze almost 14 000 ETGs extracted from the SDSS survey, in the redshift range $0.15<z<0.3$. We will show that for these old, red and passive objects the D4000 is, if the metallicity is known, an optimal indicator that correlates almost linearly with age, depending only marginally on the adopted synthetic stellar population models or star formation history (SFH). It is therefore possible to use the differential D4000 evolution of the galaxies as a tracer for their differential age evolution, and thus to set constraints to cosmological parameters in an almost model-independent way.
The paper is organized in the following way. In Sect. \ref{sec:sample}, we present the data sample, describe the different criteria with which ETGs have been selected, and give information about the D4000 definition adopted; we also discuss the mass and metallicity evaluation, and give more details about the distribution of various parameters that enhance the robustness of our selection. In Sect. \ref{sec:model0}, we describe the analysis done on the D4000-age relation obtained both using Bruzual \& Charlot (2003) and Maraston \& Str$\mathrm{\ddot{o}}$mb$\mathrm{\ddot{a}}$ck (2010) stellar population synthesis models, showing that in the range of D4000 probed by our data the linear approximation works extremely well and evaluating the slope of this relation as a function of metallicity. We therefore obtain the theoretical relation linking the D4000 with the redshift as a function of the various cosmological parameters. In Sect. \ref{sec:D4000z} we show the D4000-z relations obtained for different mass subsamples of our data, obtaining a clear redshift evolution for all the sample and a strong evidence of mass-downsizing. In Sect. \ref{sec:results} we show the results of the evaluation of $H_{0}$ and of a constant $w$, comparing them with the recent determination done by Riess et al. (2009), by the WMAP 7-years analysis and by Amanullah et al. (2010). We also discuss the robustness of our results against possible systematics from which our analysis may suffer, such as the dependence on the estimate of the absolute age of a galaxy from the $D4000_{n}$, on the star formation histories considered, on the stellar population synthesis models assumed, on the metallicity evaluation and on the not perfect assumption of the passive evolution of our ETGs, reporting at the end our estimates of $H_{0}$ and $w$ both quoting statistical and systematic errors.


\section{The sample}
\label{sec:sample}

\subsection{Early-type galaxies selection}
The sample of galaxies analyzed in this paper has been selected from the spectroscopic catalog of the Sixth Data Release of the Sloan Digital Sky Survey (SDSS-DR6). This survey provides {\it u, g, r, i, z} photometry and spectra for almost 800 000 galaxies over $\sim7000$ square degrees. Galaxies have been extracted to have a Petrosian magnitude $r<17.77$. The spectra are taken using 3-arcsec diameter fibres, positioned as close as possible to the centers of the target galaxies. The flux (and wavelenght) calibrated spectra cover the range 3800-9200 {\AA}, with a resolution of $R\sim1800$. To obtain a wider wavelength coverage for the photometry, we use a match between SDSS-DR6 galaxies and 2MASS survey provided to us by Jarle Brinchmann (private communication), with which we extend the SDSS photometry with the J, H and K bands. In this way we obtain for each galaxy a 8-bands spectral energy distribution (SED) from u to K, very important for a more robust determination of mass and of colors of the galaxy.\\
As explained before, it is fundamental to have a population of standard clocks in order to do precision cosmology with the differential age technique. In a previous paper (Moresco et al. 2010), we checked on the zCOSMOS 10k bright sample that a simple color selection is not restrictive enough to exclude starforming galaxies from the early-type galaxies selection. Following the analysis explained in that paper, we therefore decided to apply a more restrictive selection, on the basis of photometric and spectroscopic informations, to obtain a uniform, passive population of early-type galaxies (ETGs); we selected from SDSS-DR6 all galaxies matching these criteria:
\begin{itemize}
\item no strong emission lines: we decided to select only galaxies with $EW_{0}([OII]\lambda3727)<5$ {\AA} and $EW_{0}(H_{\alpha})<5$ {\AA}. These two spectroscopic features are strongly correlated with ongoing or recent episodes of star formation history. This upper limit in the equivalent width allows us to select galaxies that have experienced a limited (or null) amount of star formation in the last 1-2 Gyrs, thus choosing the most passive sample.
\item early-type SED: following the procedure explained in Zucca et al. (2006), we performed a best-fit modeling of the multi-band photometry of these galaxies using the empirical set of 62 SEDs described in Ilbert et al. (2006), where these SEDs were derived by interpolating between the four local observed spectra of Coleman et al. (1980) (from the old stellar population in M31 and M81 to Sbc, Scd, and Im SEDs) and two starburst SEDs from Kinney et al. (1996). In this way we selected only those galaxies that match a early-type SED template.
\end{itemize}
\begin{figure}[t!]
\begin{centering}
\includegraphics[angle=-90,width=0.8\textwidth]{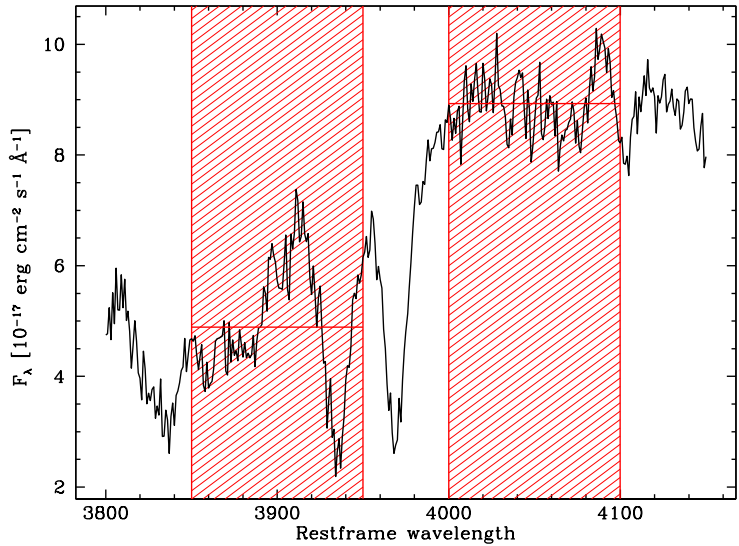}
\caption{Mean stacked spectrum of 10 ETGs randomly selected from the extreme tail of the $D4000_{n}$ distribution. The red shaded areas show the ranges used to evaluate the $D4000_{n}$ and the red horizontal lines show the mean of the flux.}
\label{fig:7}
\end{centering}
\end{figure}
For each galaxy, the 4000 {\AA} break has been taken from the MPA-JHU DR7 release of spectrum measurements ({\it http://www.mpa-garching.mpg.de/SDSS/DR7/}), which provides a complete emission line analysis for the SDSS Data Release 7 (DR7). We decided to use the D4000 definition proposed by Balogh et al. (1999), that uses narrower bands in order to be less sensitive to reddening effects (3850-3950 {\AA} and 4000-4100 {\AA}); we denote this index as $D4000_{n}$:
\begin{equation}
D4000_{n}=\frac{F_{red}}{F_{blue}}=\frac{(\lambda_{2}^{blue}-\lambda_{1}^{blue})\int_{\lambda_{1}^{red}}^{\lambda_{2}^{red}}F_{\nu}d\lambda}{(\lambda_{2}^{red}-\lambda_{1}^{red})\int_{\lambda_{1}^{blue}}^{\lambda_{2}^{blue}}F_{\nu}d\lambda}
\end{equation}
In Fig. \ref{fig:7} we plot the average stacked spectrum of 10 random galaxies, selected to be in the highest tail of the $D4000_{n}$ distribution ($D4000_{n}>2.1$) and for which the $D4000_{n}$ has been measured with high accuracy ($S/N\sim70$). The dispersion of the $D4000_{n}$ of the galaxies used for creating the stacked spectrum is $<0.01$. In the figure are shown also the ranges used to evaluate the $D4000_{n}$ and the mean fluxes in the corresponding regions.\\
As pointed out by Bernardi et al. (2006), the spectrophotometric calibration of the SDSS spectra is less reliable  around $4000$ {\AA}, due to the throughput of the spectrometer. This means that the value of the $D4000_n$ can be affected by this problem for galaxies in the lowest redshift range (Bernardi, private communication). In order to avoid this potential source of uncertainty, we decided to select our final sample at $z>0.15$. Moreover, since the number of galaxies decreases very rapidly at $z>0.3$, we excluded galaxies beyond z=0.3. Thus, the final redshift range of our sample is $0.15<z<0.3$.\\
As explained in the forthcoming section, the metallicity estimate for these galaxies has been taken from the SDSS-DR4 analysis performed by Gallazzi et al. (2005). After the match between our sample and the galaxies for which the metallicity estimate is publicly available, we end up with a sample of 13 987 galaxies, for each one having a wide photometry coverage (u to K), spectroscopic redshift, spectral index measures ($D4000_{n}$, $EW_{0}(H\alpha)$ and $EW_{0}([OII]\lambda3727)$), metallicity and mass estimates.

\subsection{Mass, metallicity and star formation history estimation}
\label{sec:massmet}
\begin{figure}[t!]
\subfigure{
\mbox{
\includegraphics[angle=0,width=0.49\textwidth]{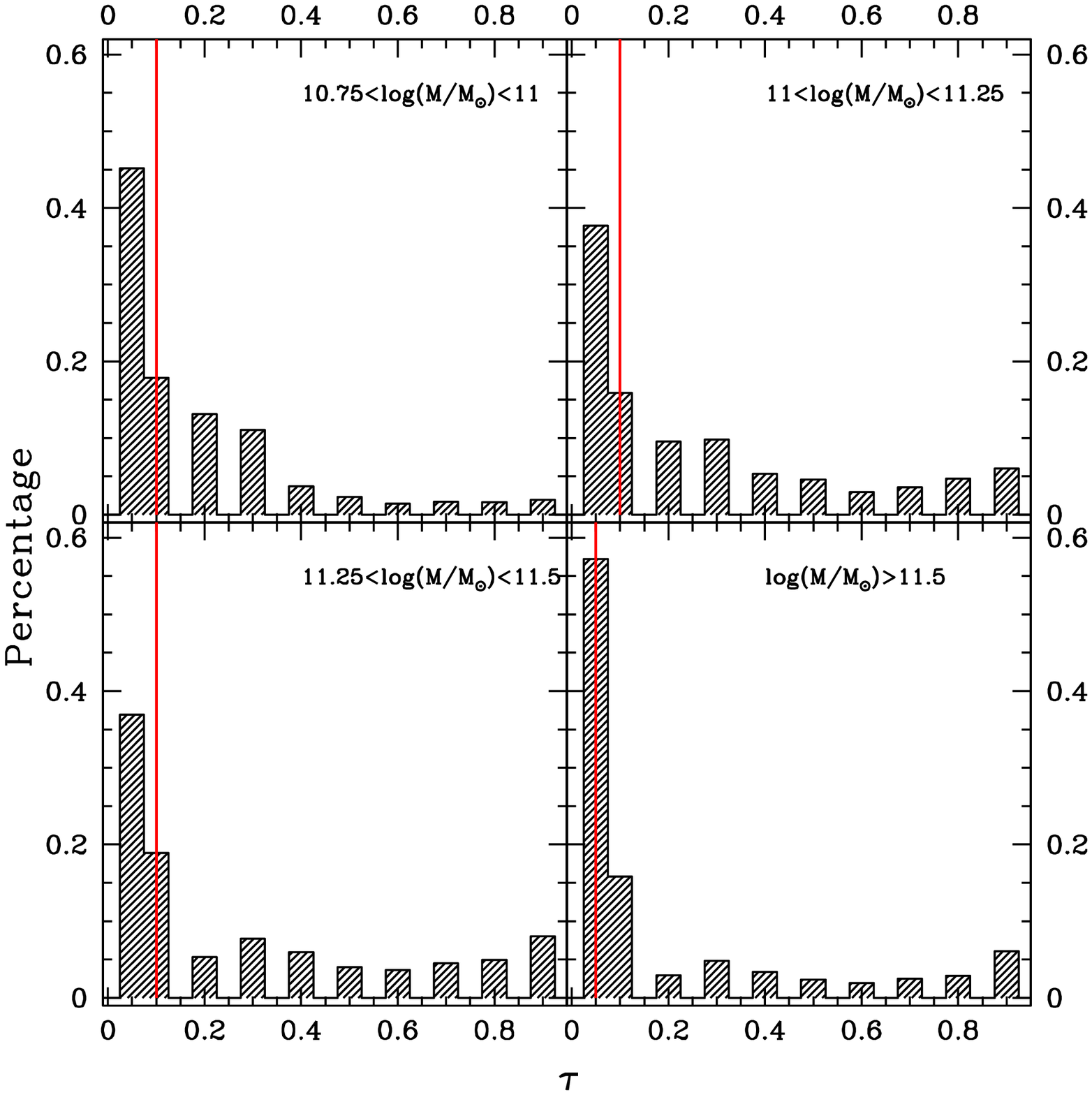}
\includegraphics[angle=0,width=0.49\textwidth]{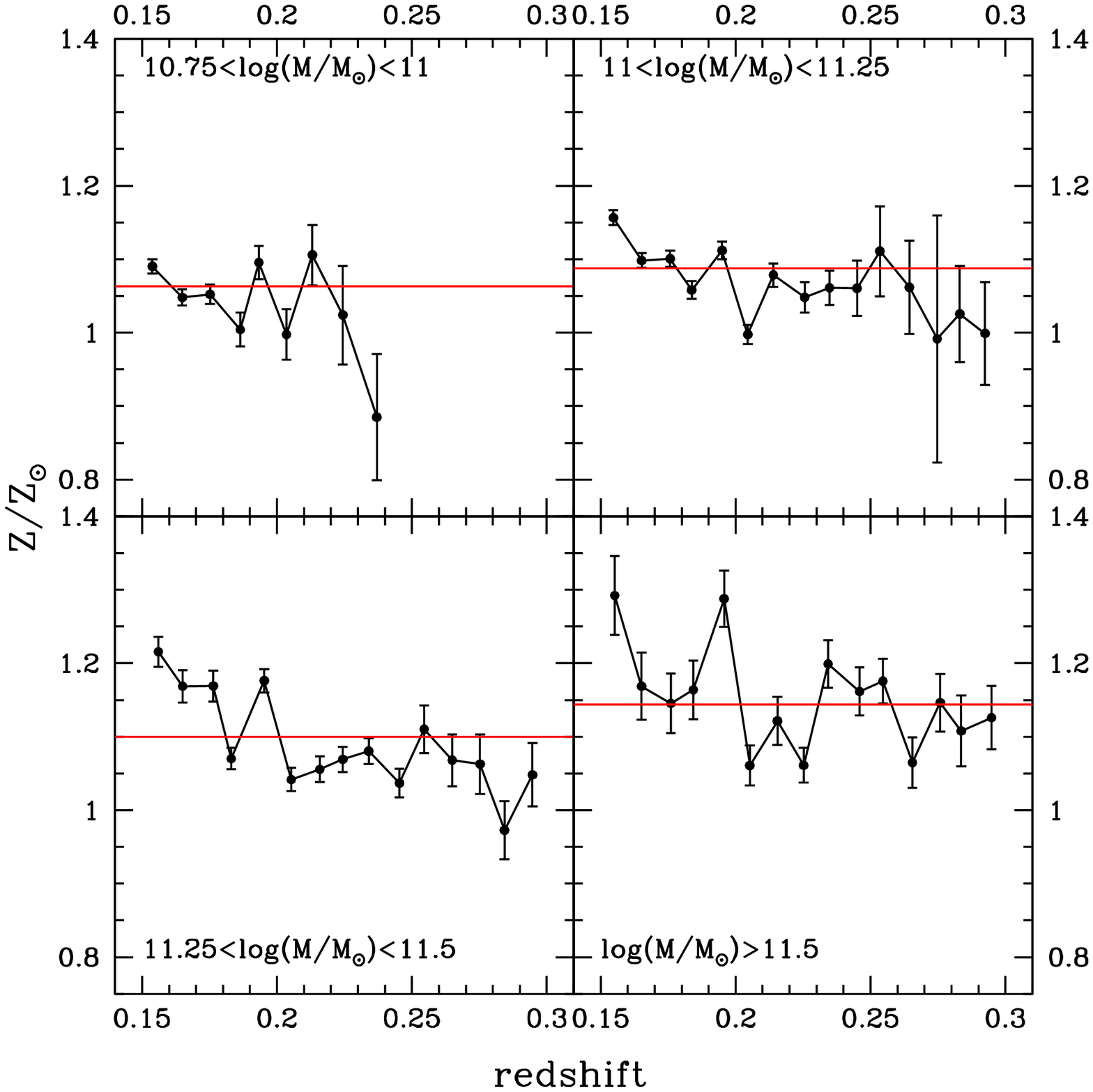}
}
}
\subfigure{
\mbox{
\includegraphics[angle=0,width=0.49\textwidth]{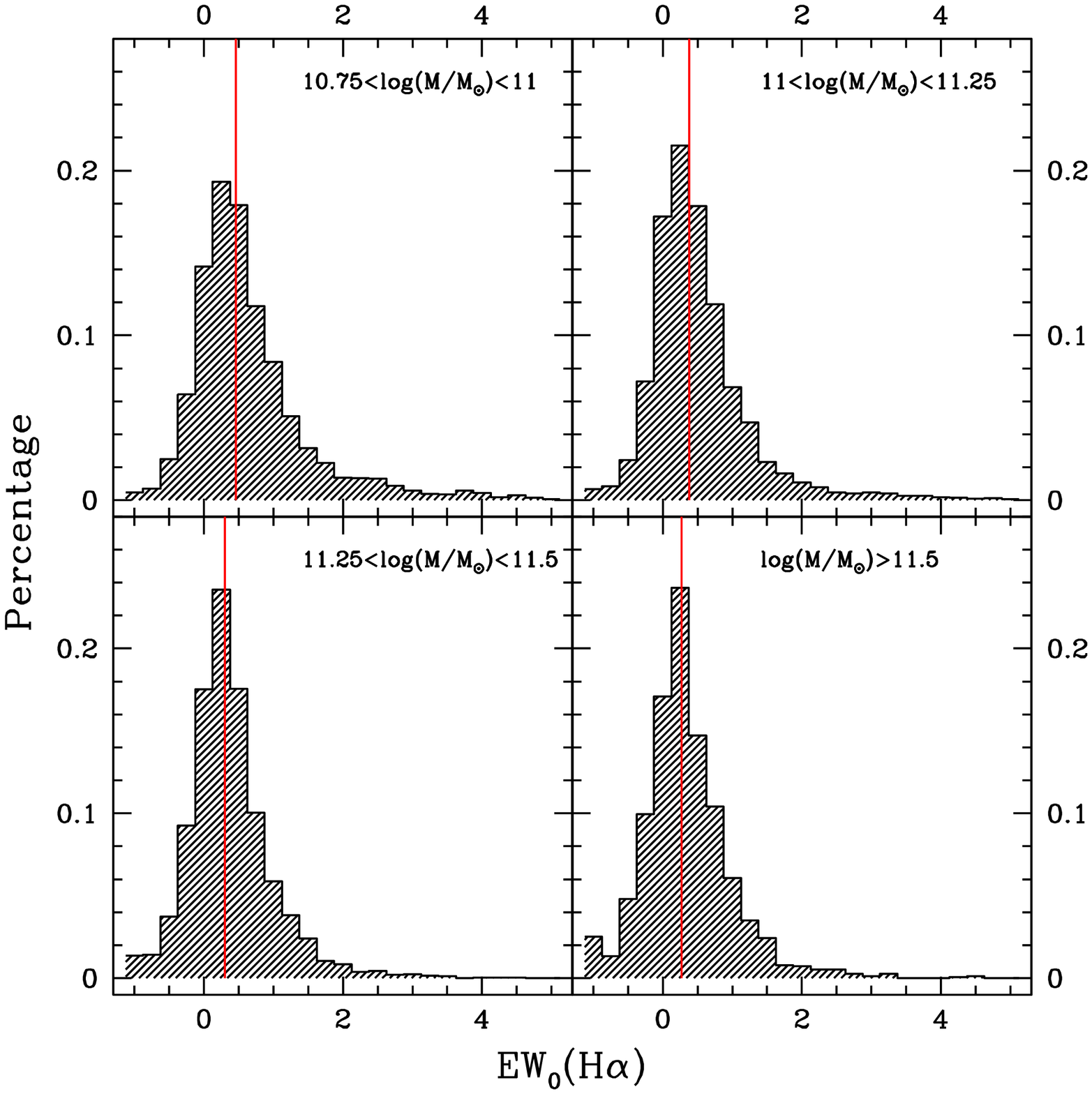}
\includegraphics[angle=0,width=0.49\textwidth]{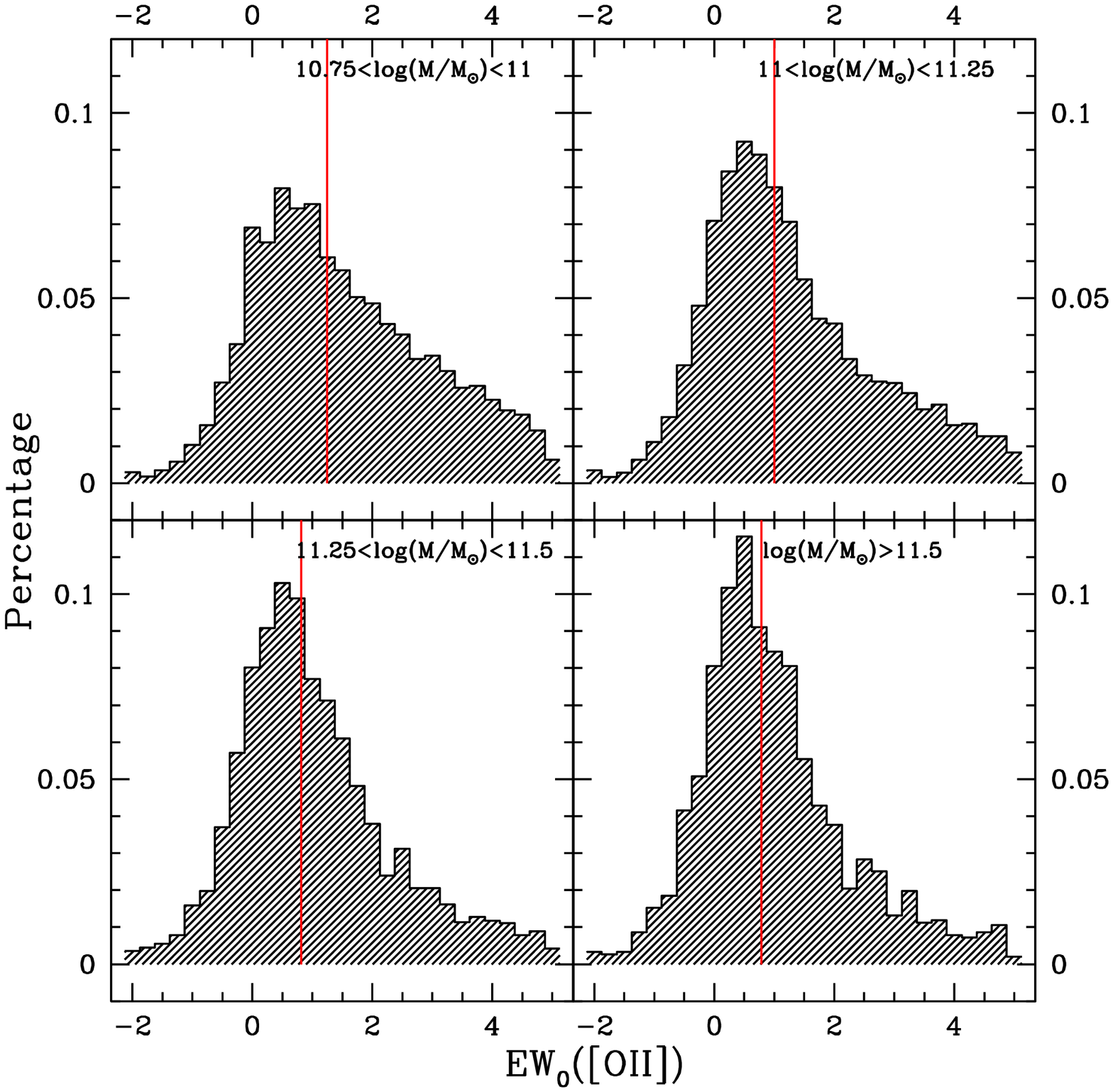}
}
}
\caption{In the upper panels, distributions of $\tau$ values (left panel) and metallicity-redshift relations (right panel) in different mass subsamples. In the lower panels, distributions of the $EW_{0}(H\alpha)$ (left panel) and $EW_{0}([OII]\lambda3727)$ (right panel). The red lines represent the median of the distributions.}
\label{fig:6}
\end{figure}
\begin{figure}[t!]
\begin{centering}
\includegraphics[angle=-90,width=0.8\textwidth]{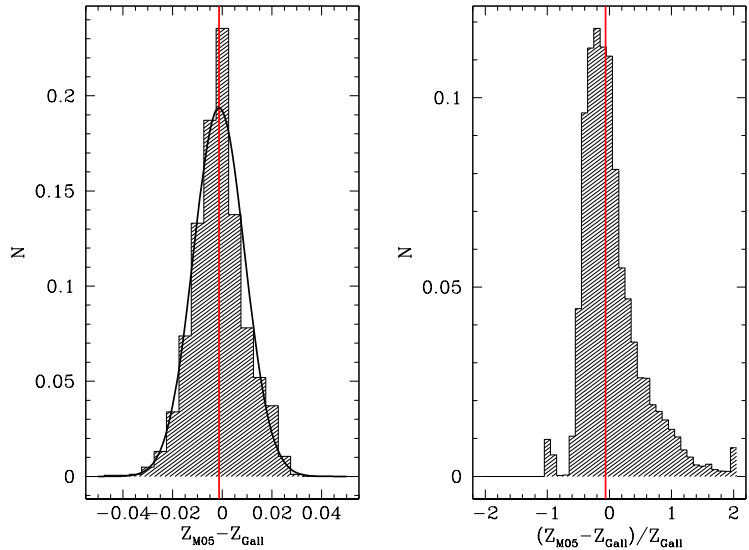}
\caption{Distributions of the difference $(Z_{Gallazzi}-Z_{VESPA})$ (left panel) and of the percentage difference $(Z_{Gallazzi}-Z_{VESPA})/Z_{Gallazzi}$.}
\label{fig:10}
\end{centering}
\end{figure}

To study the properties of our sample of ETGs, we used both photometric and spectroscopic information. To retrieve information about the metallicities of those galaxies, we used the analysis performed by Gallazzi et al. (2005), in which they derived stellar metallicities for a sample of SDSS-DR2, afterward extended to SDSS-DR4 ({\it www.mpa-garching.mpg.de/SDSS/DR4/Data/stellarmet.html}). Their constraints are set by the simultaneous fit of five spectral absorption features (D4000, $H\beta$ and $H\delta_{a}+H\gamma_{a}$ as age-sensitive indices and $[Mg_{2}Fe]$ and $[MgFe]'$ as metal-sensitive indices, all of which depend negligibly on the $\alpha/Fe$ ratio), which are well reproduced by Charlot \& Bruzual (2003, BC03) population synthesis models (see Gallazzi et al. 2005 for further details).\\
To verify that the metallicity estimation is not biased by the choice of the particular star formation history done by Gallazzi et al. (2005), we decided to compare their estimates with the ones done with VESPA code (Tojeiro et al. 2007). The approach of VESPA is to recover robust star formation and metallicities histories using the full spectral range of a galaxy from synthetic models; in order to estimate those quantities, it allows for a completely free-form star formation, imposing no prior on it. This code has been applied to the entire SDSS-DR7, obtaining a catalog of stellar masses, detailed star formation and metallicity histories (Tojeiro et al. 2009, {\it www-wfau.roe.ac.uk/vespa/}).\\
Despite the completely different assumption on the star formation histories, the distribution of the difference \mbox{$(Z_{Gallazzi}-Z_{VESPA})$} is a gaussian with mean value -0.01 and dispersion 0.1; the distribution of the percentage difference \mbox{$(Z_{Gallazzi}-Z_{VESPA})/Z_{Gallazzi}$} has a mean value of $4.8\%$, with a $51.3\%$ dispersion. So there exists no bias between the two metallicity evaluations, and this analysis ensures the reliability of Gallazzi et al. (2005) metallicity estimates (see Fig. \ref{fig:10}).

The masses are estimated by performing a best-fit to the multicolor spectral energy distribution (SED), using the observed magnitudes in 8 photometric bands from $u$ to $K$; to fit our data, we decided to use a grid of models from BC03 stellar population models. The grid has been created with delayed exponential SFH with $SFR(t,\tau)\propto(t/\tau^{2})exp(-t/\tau)$ (where $\tau$ is chosen in the range $0.05\leq\tau\leq1$ Gyrs), reddening in the range $0<A_{V}<1$ and free ages (see Pozzetti et al. 2010 for more details). Many recent works on early-type galaxies evolution has shown clearly the existence of a mass-downsizing effect, i.e. more massive galaxies have assembled their mass before less massive ones. Cimatti et al. (2006) analyzed a sample of COMBO-17 and DEEP2 ETGs, finding that the amount of evolution for the ETG population depends critically on the range of luminosity and masses considered. Scarlata et al. (2007b) show similar results for the photometric survey COSMOS, using both morphologically and photometrically selected
subsamples of early-type galaxies. Pozzetti et al. (2010), studying galaxies obtained by the zCOSMOS survey confirm a mass dependent evolution of galaxies up to $z\sim1$ in particular for ETGs, i.e. galaxy number density increases with cosmic time faster for less massive galaxies and $<20\%$ for massive ones. Ilbert et al. (2010), extend the result at $z>1$, selecting ETGs in SCOSMOS (Sanders et al. 2007) up to $z_{\rm photo}=2$, finding a more rapid evolution for massive galaxies at higher redshifts (by a factor of 15-20 between $z=1.5-2$ and $z=0.8-1$). The analysis performed by Thomas et al. (2005, 2010) on a sample taken from the SDSS survey reveals that galaxies above $log(M/M_{\odot})\sim12$ have completed to assemble their mass already at $z\approx3$, while this process happens only around $z\approx1$ for galaxies with $log(M/M_{\odot})\sim10$. Moresco et al. (2010) confirmed this result analyzing a sample of ETGs extracted from the zCOSMOS 10k bright spectroscopic sample (see Lilly et al. 2009 for further details on the parent sample); they found that the mean redshift of formation for galaxies with $log(M/M_{\odot})\sim11$ is around $z\approx2$, while for galaxies with $log(M/M_{\odot})\sim10$ it is around $z\approx1$.\\
In order to select the oldest population from our ETGs sample, we therefore decided to select the high-mass tail of the mass distribution, considering $log(M/M_{\odot})>10.75$. Moreover, to avoid possible bias from mass-downsizing effect, we decided to split our sample in narrow mass bins, to have a more homogenous sampling of redshift of formation and metallicity for these galaxies. We divided our sample in four mass bins, with $10.75<log(M/M_{\odot})<11$, $11<log(M/M_{\odot})<11.25$, $11.25<log(M/M_{\odot})<11.5$, and $log(M/M_{\odot})>11.5$. The division our ETGs into four subsamples of the width of $\Delta log(M/M_{\odot})=0.25$ has been done to keep the statistical significance of the subsamples high enough, having for each subsample $N_{gal}\gtrsim 3500$ (except for the highest mass bin, for which $N_{gal}\sim1500$); we adopted the cut $log(M/M_{\odot})>10.75$ in order both to select the most massive galaxies of our sample and to have four mass bins with a sufficient number of galaxies to perform a reasonable statistical analysis. We checked that the difference of the median mass along the redshift range is 0.11 dex on average and that the median metallicity is constant in the considered range redshift within a 5\% percent level, on average (see upper-right plot of Fig. \ref{fig:6}).
\begin{table}[b!]
\begin{center}
\small
\begin{tabular}{cccccc}
\hline \hline
mass range & median mass & median $Z/Z_{\odot}$ & median $EW_{0}(H\alpha)$ & median $EW_{0}([OII])$&\# galaxies\\
$log(M/M_{\odot})$&$log(M/M_{\odot})$&$Z/Z_{\odot}$&[\AA]&[\AA]\\
\hline
10.75-11&$10.9\pm0.02$&$1.061\pm0.006$&$0.463\pm0.009$&$1.24\pm0.03$&3452\\
11-11.25&$11.12\pm0.03$&$1.088\pm0.004$&$0.376\pm0.007$&$1\pm0.02$&5429\\
11.25-11.5&$11.25\pm0.02$&$1.1\pm0.006$&$0.305\pm0.007$&$0.81\pm0.02$&3591\\
$>11.5$&$11.63\pm0.04$&$1.144\pm0.009$&$0.27\pm0.01$&$0.78\pm0.03$&1515\\
\hline \hline
\end{tabular}
\normalsize
\caption{Median value of the mass (in logarithmic units), of the metallicity (in solar metallicity units), and of the $H\alpha$ and $[OII]\lambda3727$ equivalent widths of the ETGs in different mass subsamples. We also quote for each bin the number of galaxies.} \label{tab:1}
\end{center}
\end{table}

The $\tau$ parameter of the SFH obtained from the best-fit to the SEDs of the galaxies shows a distribution characteristic of passive galaxies, with median values below 0.2 Gyrs for all the mass subsamples (see upper-left plot of Fig. \ref{fig:6}). This result is in agreement with many ETGs analysis (Cimatti et al. 2008, Longhetti et al. 2008, Gobat et al. 2008), for which the majority of massive field early-type galaxies should have formed their stellar content around $z\gtrsim2$ over short (i.e. $\tau<0.1-0.3$ Gyrs) star formation time-scales.  We also studied the distributions of the Star Formation Rate (SFR) obtained from the SED-fitting, and found a median SFR equal to zero $\mathrm{M_{\odot}/yr}$ for all the mass subsamples, with a scatter of 0.11, 0.17, 0.18, 0.2 $\mathrm{M_{\odot}/yr}$, respectively for the four mass subsamples.

Since our analysis depends strongly on the selection of a sample of galaxies passively evolving, we analyzed in detail also the emission lines distributions, because they can be linked with star formation or AGN activity. We find that the median values of those distributions are $EW_{0}(H\alpha)\lesssim0.5$ {\AA} and $EW_{0}([OII]\lambda3727)\lesssim1$ {\AA}, with tails up to 2 {\AA} for the $EW_{0}(H\alpha)$ and up to 4 {\AA} for the $EW_{0}([OII]\lambda3727)$ (see \mbox{Tab. \ref{tab:1}}). The median value of the specific star formation rate, obtained from the SED-fitting, for the galaxies with a significant detection of $EW_{0}(H\alpha)$ is however null. Moreover, we verified that the $D4000_{n}$ is not significantly different between galaxies with $EW_{0}(H\alpha)<0$ and $EW_{0}(H\alpha)>0$, finding respectively $D4000_{n}=1.948\pm0.002$ and $D4000_{n}=1.944\pm0.001$.\\
We also checked the distribution of the \textsf{eClass} spectral parameter, that is a SDSS parameter quantifying the activity of a galaxy using a PCA analysis of the spectra of the galaxies; it goes from -0.35 to 0.5 for early- to late-type galaxies. For each mass subsample, it shows a distribution always below -0.1, so our selection results even stricter than the one done from Bernardi et al. (2006), in which they chose ETGs with $eClass<0$. In Fig. \ref{fig:6} we show in the upper panels the $\tau$ distribution (left panel) and the metallicity-redshift relations (right panel), while in the lower panels the distribution of the $EW_{0}(H\alpha)$ (left panel) and $EW_{0}([OII]\lambda3727)$ (right panel), each one for the four samples; the red lines represent the medians of the distributions. In Table \ref{tab:1} are reported the median values of mass, metallicity and equivalent widths of the emission lines for the different mass subsamples, with their errors; we report also in the same table the number of galaxies in each bin.

Another possible systematic trend with redshift that could, in principle, bias our analysis is the effect of using in SDSS a fixed 3 arcsec fibre, which maps different physical sizes depending on the observed redshift. However Gallazzi et al. (2005) analyzed in details the effect of this aperture bias, finding no significant difference in metallicity between high- and low-concentration galaxies (see their Sect. 3.2). We also checked that the difference in D4000 between high- and low-concentration galaxies is always $<2\%$.


\section{Linking the 4000 {\AA} break to the expansion history of the Universe}
\label{sec:model0}

\subsection{Calibration of the $D4000_{n}$-age relation}
\label{sec:model}
\begin{figure}[t!]
\begin{centering}
\mbox{
\includegraphics[angle=0,width=0.55\textwidth]{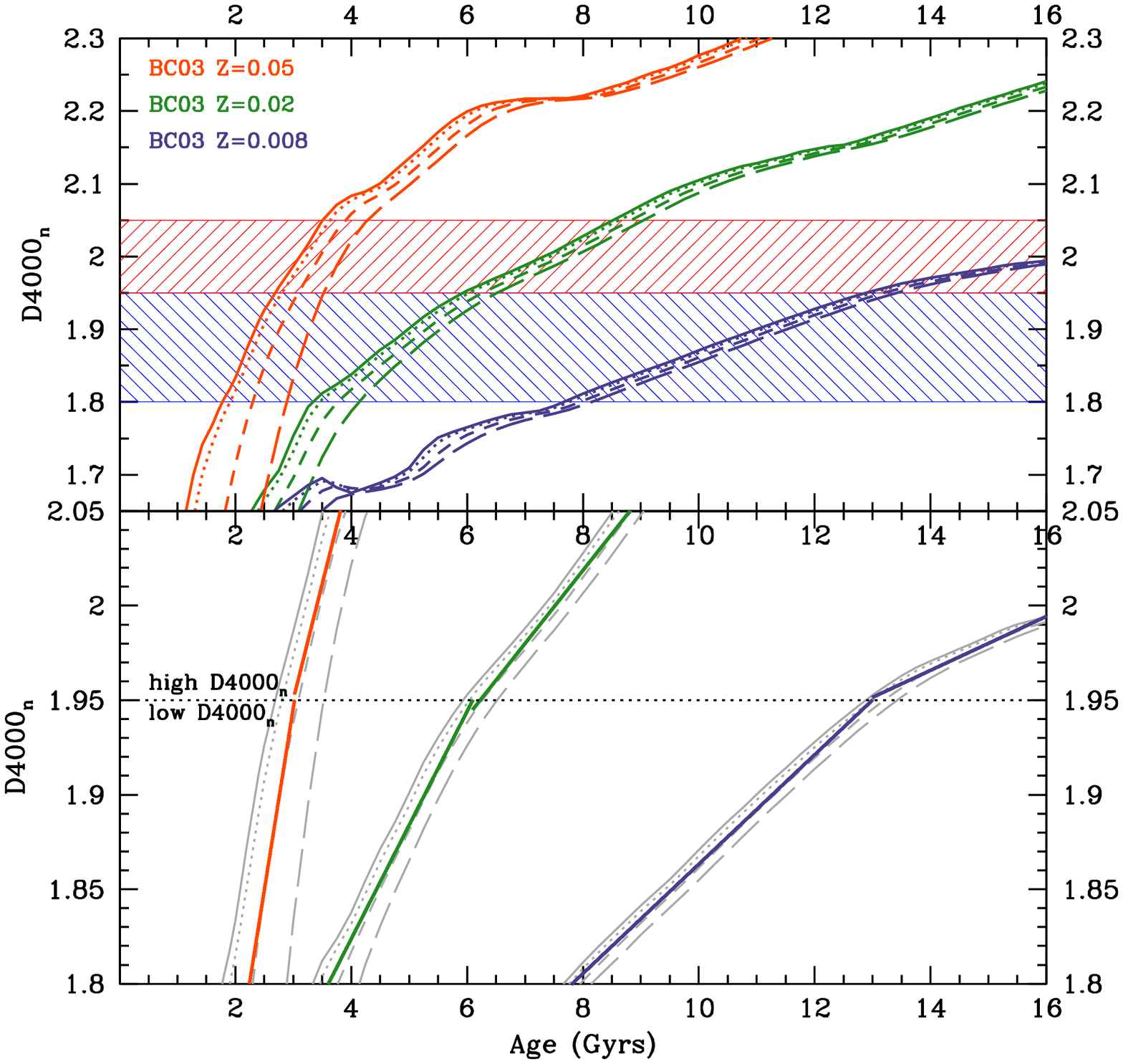}
\includegraphics[angle=0,width=0.55\textwidth]{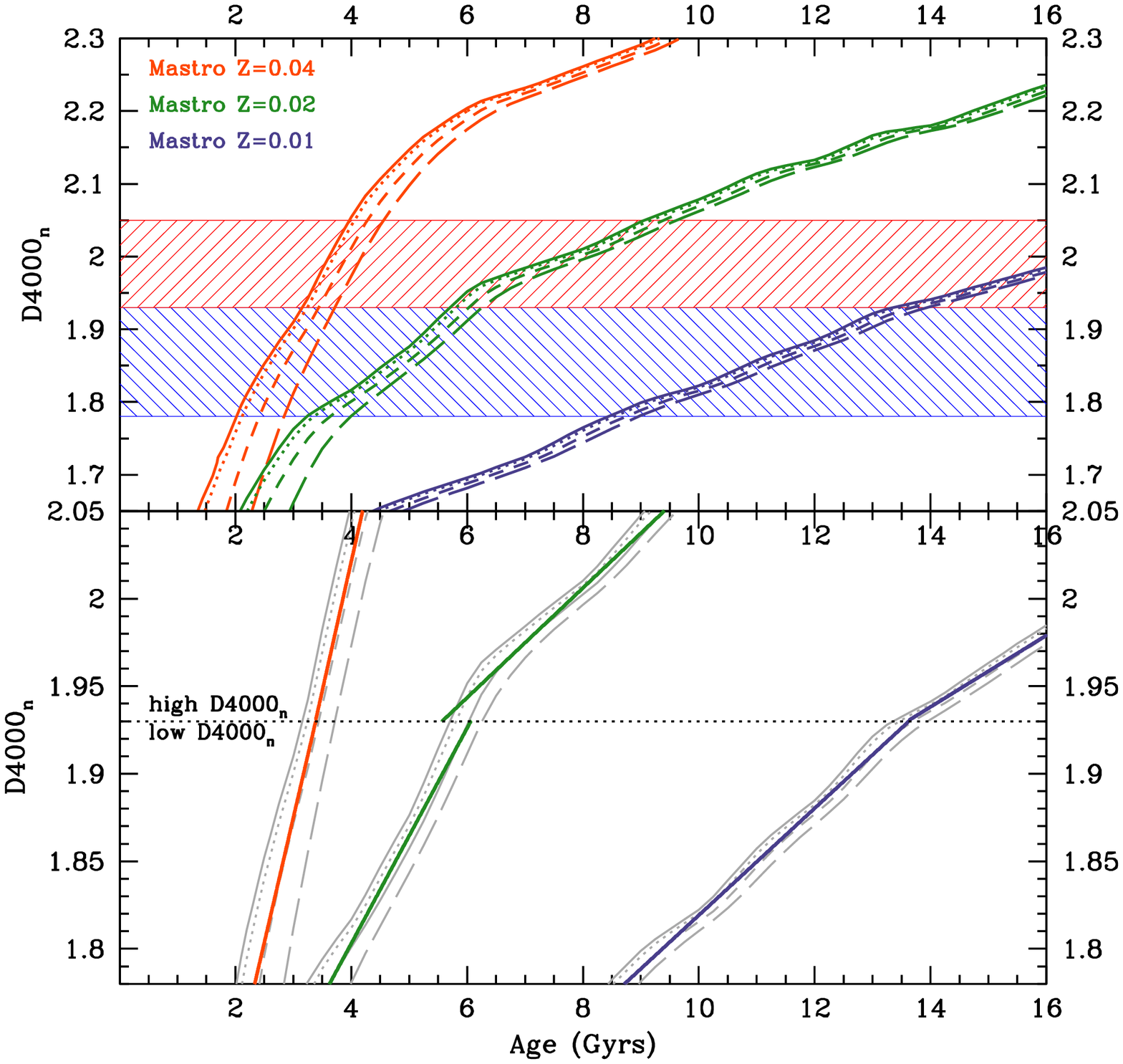}
}
\caption{$D4000_{n}$-age relation for BC03 high resolution models (left panel) and {\it Mastro} models (right panel). In the upper panels the colored lines represent models with different metallicities, with oversolar metallicity in red (respectively $Z/Z_{\odot}=2.5$ and $Z/Z_{\odot}=2$), solar in green, and undersolar in blue (respectively $Z/Z_{\odot}=0.4$ and $Z/Z_{\odot}=0.5$). The blue and the red shaded area represent the ranges of $D4000_{n}$ in the models (respectively the {\it low $D4000_{n}$} and the {\it high $D4000_{n}$} regime). The lower panel show a zoom of the interested area, where the models are shown in gray and the colored lines are the fit to the models. The dotted line shows where the change between the {\it low $D4000_{n}$} and the {\it high $D4000_{n}$} regime. It is possible to see the minor dependence of the slopes of the \mbox{$D4000_{n}$-age} relation on the SFH assumed both by visually comparing the average slopes (colored lined of the lower panels) and the slope of each single model (grey lines of the lower panels) and by looking the small errors associated to the mean values (see Table \ref{tab:2}).}
\label{fig:2}
\end{centering}
\end{figure}
In general the $D4000_{n}$ is an index that is strongly sensitive both to metallicity, star formation history and age of a stellar population. In order to break this degeneracy, we created a library of $D4000_{n}$ as a function of age using stellar population synthesis models with different metallicity, star formation history (delayed exponential SFR with $\tau=0.05, 0.1, 0.2, 0.3$ Gyrs), and ages. In order to study the possible systematics due to the choice of a particular model, we created two different libraries, one using Bruzual and Charlot (2003) stellar population synthesis models (hereafter BC03) and one using the new Maraston \& Str$\mathrm{\ddot{o}}$mb$\mathrm{\ddot{a}}$ck (2010) stellar population synthesis models (hereafter {\it Mastro}), which are based on the MILES models (Sanchez-Blazquez et al. 2006). The metallicities studied with the two models are $Z/Z_{\odot}=0.4, 1, 2.5$ for BC03 and $Z/Z_{\odot}=0.5, 1, 2$ for {\it Mastro}. The resolution of the two models is similar, 3 {\AA} across the wavelength range from 3200 {\AA} to 9500 {\AA} for BC03, and 2.3 {\AA} across the wavelength range 3525 {\AA} to 7500 {\AA} for {\it Mastro}; moreover the resolution of both models is comparable to the resolution reachable with the SDSS spectrograph, which has $R\sim1800$ between 3900-9100 {\AA}.

The choice of the grid of star formation histories and metallicities is motivated by the fact that we want to analyze ETGs from the SDSS survey: a lower value in the range of metallicity would probe $D4000_{n}$ values that are much lower than the average values found in our galaxy sample, and the SED-fitting analysis of our ETGs shows that the majority of the sample is best-fitted with SFHs with low values of $\tau$. Moreover, the choice of a passive sample strongly supports our decision of using these low $\tau$ values.

The upper panel of Fig. \ref{fig:2} shows the BC03 and {\it Mastro} $D4000_{n}$-age relations; the models with the higher metallicity ($Z/Z_{\odot}=2.5$ and $Z/Z_{\odot}=2$ respectively) are in red, the ones with $Z/Z_{\odot}=1$ in green, and the ones with the lower metallicity ($Z/Z_{\odot}=0.4$ and $Z/Z_{\odot}=0.5$ respectively) in red. For each metallicity, the models are plotted with a continuous line for $\tau=0.05$ Gyrs, with a dotted line for $\tau=0.1$ Gyrs, with a dashed line for $\tau=0.2$ Gyrs and with a long-dashed line for $\tau=0.3$ Gyrs.

We decided to study these relations in the $D4000_{n}$ range spanned by our data, roughly $1.8\lesssim D4000_{n}\lesssim2$. For both the models considered, we found that the $D4000_{n}-age$ relation presents, at each metallicity, two different slopes: one characteristic of the {\it "low $D4000_{n}$"} regime ($1.8<D4000_{n}<1.95$ for BC03 and $1.75<D4000_{n}<1.93$ for {\it Mastro}) and one characteristic of the {\it "high $D4000_{n}$"} regime ($1.95<D4000_{n}<2.05$ for BC03 and $1.93<D4000_{n}<2.05$ for {\it Mastro}). We demonstrate that, at the condition of studying the model separately in the two regimes, the linear approximation:
\begin{equation}
D4000_{n}(Z)=A(Z)\cdot age+B(Z)
\label{eq:1}
\end{equation}
is valid and accurate at fixed metallicity, having in the case of BC03 models correlation coefficients always $>0.996$ with a mean value of $0.9987\pm0.0004$ for the {\it "low $D4000_{n}$"} regime and always $>0.994$ with a mean value of $0.9976\pm0.0007$ for the {\it "high $D4000_{n}$"} regime; in the case of {\it Mastro} models the correlation coefficients are always $>0.997$ with a mean value of $0.9984\pm0.0003$ for the {\it "low $D4000_{n}$"} regime and always $>0.993$ with a mean value of $0.9986\pm0.0005$ for the {\it "high $D4000_{n}$"} regime.

The only limitation of this approximation is that it requires a direct estimate of the metallicity of the sample. However, as we will show later, this method has been proven to be robust against the metallicity evaluation, since the analysis of subsample with a median metallicity completely different give results in full agreement.
We decided to define for each metallicity a mean slope $<A>$, averaging between the slopes with same metallicity and different SFHs. The mean values of the slopes are reported in Table \ref{tab:2}. The use of an average slope instead of the single values is motivated by the fact that the dependence on the SFH is much less significant than the dependence on age and metallicity (see upper panels of Fig. \ref{fig:2}), as testified by the small errors associated to the mean values (see Table \ref{tab:2}). In the lower panels of Fig. \ref{fig:2} we plot in gray the models, as explained before, while the colored lines represent the average linear fit to those models; the comparison between the colored and the grey lines shows visually the similarity between the average slopes and the slope of each single model. The dotted horizontal line represent the separation between the {\it "low $D4000_{n}$"} and the {\it "high $D4000_{n}$"} regime. From this figure are clearly evident the different slopes present in the two range of $D4000_{n}$, for each metallicity.
\begin{table}[b!]
\begin{center}
\begin{tabular}{cccc}
\hline \hline
& & low $D4000_{n}$ & high $D4000_{n}$\\
\hline
BC03 & $<A>(Z/Z_{\odot}=0.4)$&$0.02893\pm0.00002$&$0.0159\pm0.0003$\\
BC03 & $<A>(Z/Z_{\odot}=1)$&$0.0601\pm 0.0006$&$0.0359\pm0.0005$\\
BC03 & $<A>(Z/Z_{\odot}=2.5)$&$0.1926\pm0.0113$&$0.1223\pm0.0013$\\
\hline
{\it Mastro} & $<A>(Z/Z_{\odot}=0.5)$&$0.03136\pm0.0004$&$0.02427\pm0.00005$\\
{\it Mastro} & $<A>(Z/Z_{\odot}=1)$&$0.0582\pm 0.002$&$0.0341\pm0.0005$\\
{\it Mastro} & $<A>(Z/Z_{\odot}=2)$&$0.135\pm0.0083$&$0.1257\pm0.004$\\
\hline \hline
\end{tabular}
\caption{Mean slopes $<A>$ of the $D4000_{n}$-age relation (see Eq. \ref{eq:1}) for the BC03 and {\it Mastro} models with different metallicity and for the two $D4000_{n}$ regimes.}
\label{tab:2}
\end{center}
\end{table}

Since, as Eq. \ref{eq:1} shows, the slope $A$ represents the conversion parameter between the $D4000_{n}$ and the age of a galaxy, a fundamental step of our approach is to assign to correct $A$ value to the metallicity evaluated in each sample. Therefore we decided to interpolate the values reported in \mbox{Tab. \ref{tab:2}} with their errors with a quadratic function, obtaining a $<A>$-metallicity relation; in this way it is possible to find the correct $<A>$ parameter considering the metallicities of each mass subsamplesusing this relation, and they are given in \mbox{Tab. \ref{tab:1}}.\\
In our approach we take into account also the effect of averaging the slope $A$ between different star formation histories, since the quadratic function has been defined with a best fit to the $<A>$-metallicity relation, where the errors in the parameter $<A>$ quantify exactly the difference of the slope between the different star formation histories considered. The size of the errors by themselves shows that the dependence of the slope on the star formation history is small. We find that the percentage difference between models with different SFHs, averaged in the metallicity range spanned by our data, is $2.8\%$ in the {\it "low $D4000_{n}$"} regime and $3.1\%$ in the {\it "high $D4000_{n}$"} regime for BC03 models and $12.4\%$ in the {\it "low $D4000_{n}$"} regime and $3.9\%$ in the {\it "high $D4000_{n}$"} regime for {\it Mastro} models. We decided to take into account this difference in our analysis, adding these errors in the cosmological parameter evaluation.

This analysis has been carried out also with the new {\it Mastro} models to study the dependence of the cosmological parameter evaluation on a particular library of models. By comparing  the plots from Fig. \ref{fig:2}, it can be found that the $D4000_{n}$-age relations have a different normalization, yielding the {\it Mastro} models to younger ages on average. However, even with different absolute $D4000_{n}$ values, the study of the slope of the relations gives values in good agreement with the ones obtained from the BC03 models.\\
This results is of key importance in our approach. On one hand it is fundamental to remind that our analysis is based only on the {\it relative} change of $D4000_{n}$, and not on its {\it absolute} value. Thus, the variation in the normalization between the different models is irrelevant. On the other hand, the fact that the slope of the $D4000_{n}$-age relation remains almost unchanged passing from BC03 to {\it Mastro} models demonstrate the robustness of our method against the choice of different stellar population synthesis models. In Section \ref{sec:syst} we will quantify what is the effective consequence of this difference in the parameter estimation.

\subsection{A linear model}
\label{sec:theormodel}
The cosmological expansion history is described by the Hubble parameter evolution:
\begin{equation}
H(z)=-\frac{1}{1+z}\frac{dz}{dt}
\label{eq:2}
\end{equation}
The analysis of the differential age evolution of ETGs, that trace the differential age evolution of the Universe, fully determine H(z), and therefore the cosmological parameters.\\
As shown in section \ref{sec:model}, there exists, in the range of $D4000_{n}$ probed by our data and at fixed metallicity Z, a linear relation between the $D4000_{n}$ and the age of a galaxy; therefore, if we use Eq. \ref{eq:1}, Eq. \ref{eq:2} can be easily rewritten as a function of the differential evolution of the $D4000_{n}$ with a proper conversion parameter, where this conversion parameter is given by the slope A of the $D4000_{n}$-age relation:
\begin{equation}
H(z)=-\frac{A}{1+z}\frac{dz}{dD4000_{n}}
\label{eq:3}
\end{equation}
This approach, provided that the galaxies are passive and evolved passively, is promising because:
\begin{itemize}
\item it relies on a direct observable of a galaxy, that can be easily measured from spectra;
\item it depends only on the slope of the $D4000_{n}$-age relation, and not on its overall normalization;
\item it depends only marginally on the SFH assumptions;
\item it is not influenced by dust extinction (Hamilton 1985, Balogh et al. 1999).
\end{itemize}
Equation \ref{eq:3} can be reformulated in order to make explicit the dependence on the various cosmological parameters in the following way:
\begin{eqnarray}
\displaystyle
D4000_{n}(z)&=&\frac{A}{H_{0}}\int_{0}^{\frac{1}{1+z}}y^{1/2}\left\{\Omega_{m}+\Omega_{DE}\left[y^{-3(w_{0}+w_{a})}e^{3w_{a}(y-1)}\right]\right\}^{-1/2}dy\nonumber\\
&&+constant
\label{eq:4}
\end{eqnarray}
where we have used the standard parametrization $w(z)=w_{0}+(\frac{z}{1+z})w_{a}$.


\section{Deriving the $D4000_{n}$-redshift relation}
\label{sec:D4000z}
\begin{figure}[t!]
\begin{centering}
\includegraphics[angle=-90,width=0.76\textwidth]{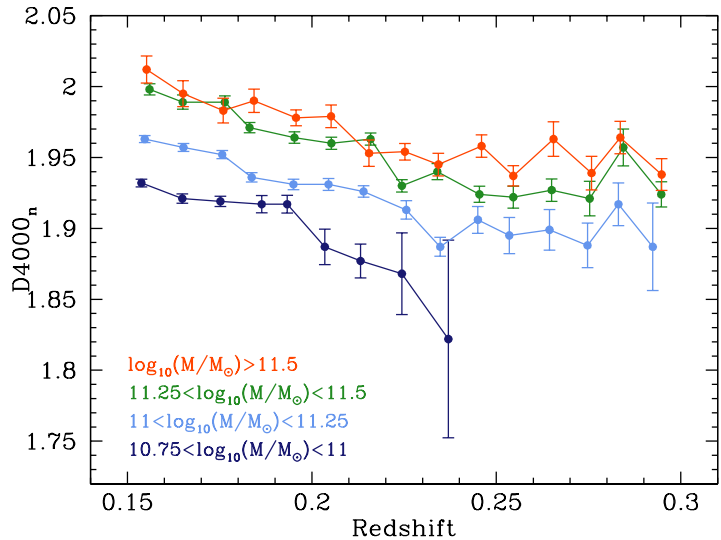}
\caption{$D4000_{n}$-redshift relation for ETGs in different mass subsamples.}
\label{fig:1}
\end{centering}
\end{figure}

For each mass subsample, we evaluated the median $D4000_{n}$ in narrow redshift bins ($\Delta z=0.01$); the associated errors are standard errors on the median.\footnote[1]{The error on the median are evaluated as the median absolute deviation/sqrt(N), where the {\it "median absolute deviation"} (MAD) $MAD=1.482*median(|D4000_{n}-median(D4000_{n})|)$. (see Hoaglin et al. 1983)}
Our results are shown in Fig. \ref{fig:1}. In the lower mass range, we limit our analysis to $z\lesssim0.24$, because, given the magnitude limit of our sample, above $z\approx0.24$ we have only $N_{gal}=9$.\\
It is impressive that for each mass subsample we find a clear $D4000_{n}$-z relation. Since we have proven in Sect. \ref{sec:massmet} that the metallicity evolution with redshift is negligible within each mass subsample, this directly witnesses the differential age evolution of ETGs, making evident that galaxies at low redshift have always higher break with respect to galaxies at high redshift. Moreover Fig. \ref{fig:1} provide also good evidence of mass-downsizing, showing that more massive galaxies present at each redshift higher $D4000_{n}$ breaks with respect to less massive ones, hence having higher ages, since the difference in metallicity is small.

Comparing our $D4000_{n}$ values shown in Fig. \ref{fig:1} with theoretical values, we find that with these metallicities the relative age evolution of the ETGs is $\approx1.2$ Gyrs in the range $0.15<z<0.3$, fully compatible with the theoretical age evolution expected in a $\Lambda$CDM Universe within this redshift range.\\
The absolute ages of these ETGs go on average from $\approx5-6$ Gyrs in the lowest mass bin to $\approx6-7$ Gyrs in the highest mass bin, so it seems that there are some problems of normalization between the ages obtained from the $D4000_{n}$ and the expected absolute ages, that are around $\approx8-10$ Gyrs for ETGs in this redshift range. However on the one hand, it is fundamental to remember again that our approach relies completely on the differential $D4000_{n}$ evolution of ETGs, so in the cosmological parameter evaluation we are insensitive to the overall normalization. On the other hand, it is also important to stress that we are using median values, so we are averaging between different galaxies ages; if we consider only the upper envelope of the $D4000_{n}$ distribution, which represents the oldest population of ETGs, we retrieve again absolute ages of $\approx8$ Gyrs.\\
Another crucial point in favor of using our approach is that the use of this linear conversion from $D4000_{n}$ to the age of a galaxy keeps by definition unchanged the slope of the redshift dependence, allowing a better estimate of cosmological parameters.

As stressed before, to apply our technique it is fundamental to obtain the correct A parameter, that is the conversion factor from the $D4000_{n}$ to the age of a galaxy. Therefore we need to associate the proper metallicity to each mass subsample. Since, as already underlined, we verified that the metallicity evolution with redshift is negligible within each subsample, we studied the medians of the metallicity distribution, with their standard errors. As predicted by Gallazzi et al. (2006), we find that all ETGs have slightly oversolar metallicity values, increasing as a function of mass.\\
We use the $<A>$-metallicity relation defined in Sect. \ref{sec:model} to find the proper $<A>$ parameter corresponding to the metallicity of each mass subsample, giving the median metallicity reported in Table \ref{tab:1}. We decided to use the values corresponding to the {\it "low $D4000_{n}$"} regime for the lower mass bins ($10.75<log(M/M_{\odot})<11$, $11<log(M/M_{\odot})<11.25$, and $11.25<log(M/M_{\odot})<11.5$), and the values corresponding to the {\it "high $D4000_{n}$"} regime for the highest mass bin ($log(M/M_{\odot})>11.5$).


\section{Constraints on the cosmological parameters}
\label{sec:results}
\begin{figure}[t!]
\begin{centering}
\includegraphics[angle=-90,width=0.8\textwidth]{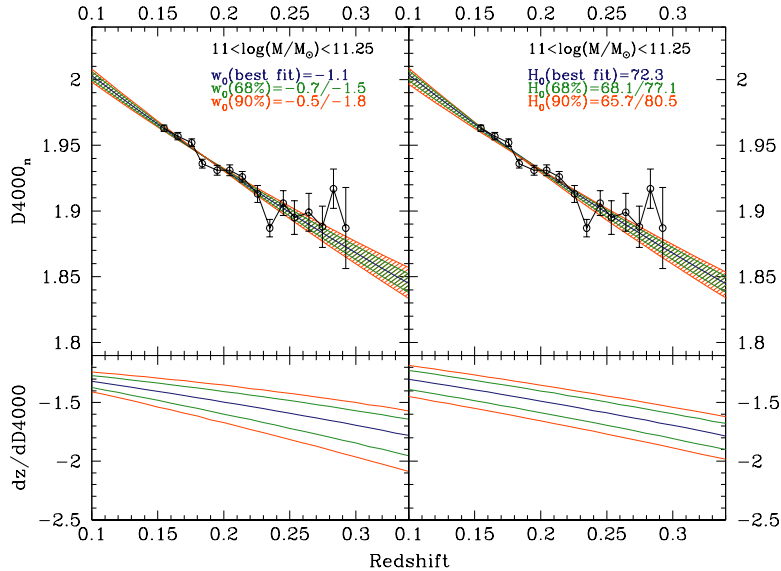}
\caption{$D4000_{n}$-redshift relation (upper panels) and $dz/dD4000_{n}$ relation for different theoretical lines. in the upper panels the black points are the relation obtained for the $11<log{M/M_{\odot}}<11.25$ mass subsample. The blue lines represent the best-fit to the data, using $\Omega_{M}=0.27$, $\Omega_{DE}=0.73$, $H_{0}=74.2$ in the left panels and $\Omega_{M}=0.27$ and $\Omega_{DE}=0.73$ in the right panels. The green and red lines show respectively the curves representing the $68\%$ and $90\%$ errors to the best-fit.}
\label{fig:8}
\end{centering}
\end{figure}
The $D4000_{n}$-z relations obtained from our selection of ETGs extracted from the \mbox{SDSS-DR4} are then used to set constraints on the Hubble constant $H_{0}$ and on the dark energy equation of state parameter $w$, assuming $w=\mathrm{constant}$ ($w=w_{0}$, $w_{a}=0$).

The most recent evaluation of the Hubble parameter has been derived by Riess et al. (2009) by analyzing the magnitude-redshift relation of 240 low-z Type Ia supernovae at $z < 0.1$. The absolute magnitudes of supernovae are calibrated using new observations from HST of 240 Cepheid variables in six local Type Ia supernovae host galaxies and the maser galaxy NGC 4258. This refurbished distance ladder based on extensive use of differential measurements allowed them to obtain $H_{0}=74.2\pm3.6\;km\,s^{-1} Mpc^{-1}$, including both statistical and systematic errors.\\
This measurement has been proved to be extremely useful also for a new determination of the dark energy equation of state parameter obtained by the seven-year WMAP observation (Komatsu et al. 2010): by combining the WMAP data with the latest distance measurements from the Baryon Acoustic Oscillations (BAO) in the distribution of galaxies (Percival et al. 2010) and the Hubble constant ($H_{0}$) measurement (Riess et al. 2009) it was possible to set the constraint \mbox{$w = -1.10\pm0.14$} (68\% CL).\\
The most recent estimate of $w$ comes from the analysis of a sample of 557 SNe called Union2 compilation, obtained combining the Union compilation with the HST-light curves and spectra of six new high redshift SNe Ia (Amanullah et al. 2010); their analysis gives results in full agreement with WMAP estimate, with a $w=-0.997\pm0.08$ for a flat $\mathrm{\Lambda CDM}$ Universe and $w=-1.038\pm0.09$ for a $\mathrm{\Lambda CDM}$ Universe with curvature, both assuming $w=$constant.

In the next sections we will show the strength of our approach in determining both $H_{0}$ and $w$, by analyzing individually the different mass subsample and then performing a joint analysis, unifying the constraints derived from the single measurements.\\
As an example, we show in Figure \ref{fig:8} in black the $D4000_{n}-z$ relation for the mass bin $11<log{M/M_{\odot}}<11.25$. We performed a best-fit to this relation, in order to obtain $w$ (in the left plot) and $H_{0}$ (in the right plot). The blue curves show the best-fits to the data; we also plot in green and red the curves representing the $68\%$ and $90\%$ errors to the best-fits, respectively. In the upper-left panel, the best-fit model has been evaluated at fixed $\Omega_{M}=0.27$, $\Omega_{DE}=0.73$ (as obtained by WMAP 7-years analysis), $H_{0}=74.2$ (as obtained by Riess et al. 2009), to set constraint on $w=\mathrm{constant}$; in the upper-right panel we set $w=-1$, $\Omega_{M}=0.27$, $\Omega_{DE}=0.73$, and search for the best-fit $H_{0}$ value. The size of the errors of our data makes clear the strength of the differential age approach to set constraints on the cosmological parameters. In the lower panels we show the differential evolution of the $D4000_{n}$, $dz/dD4000_{n}$, showing how much these relations are sensitive to the variation of $H_{0}$ and $w$.

\subsection{Statistical methods}
\label{sec:stat}
We fit the $D4000_{n}-z$ relations shown in Fig. \ref{fig:1} with the formula given in Eq. \ref{eq:3}, using a standard $\chi^{2}$ approach. The theoretical $D4000_{n}(z)$ relation is a function of the $A$ parameter and of various cosmological parameters ($H_{0}$, $\Omega_{m}$, $\Omega_{DE}$, $w$, $w_{a}$). Therefore we will have $\chi^{2}(\alpha_{1}, \alpha_{2}, ..., \alpha_{N})$, where $\alpha_{i}$ are the free parameter of the fit. The associated likelihood is given by $\mathcal{L}(\alpha_{1}, \alpha_{2}, ..., \alpha_{N})\propto e^{-(\chi^{2}(\alpha_{1}, \alpha_{2}, ..., \alpha_{N}))/2}$. Being interested only in the evaluation of one parameter, we decided to marginalize this likelihood over the parameters we are not interested, and we obtain:
\begin{equation}
\mathcal{L}(\alpha_{i})=\int e^{-\frac{\chi^{2}(\alpha_{1}, \alpha_{2}, ..., \alpha_{N})}{2}} d\alpha_{1}... d\alpha_{i-1}d\alpha_{i+1}...d\alpha_{N}
\end{equation}
In the following sections we will explain the adopted priors and assumptions to estimate the Hubble constant $H_{0}$ and the dark energy equation of state parameter $w$.

\subsection{The Hubble constant}
\label{sec:H0}
\begin{figure}[t!]
\mbox{
\includegraphics[angle=0,width=0.49\textwidth]{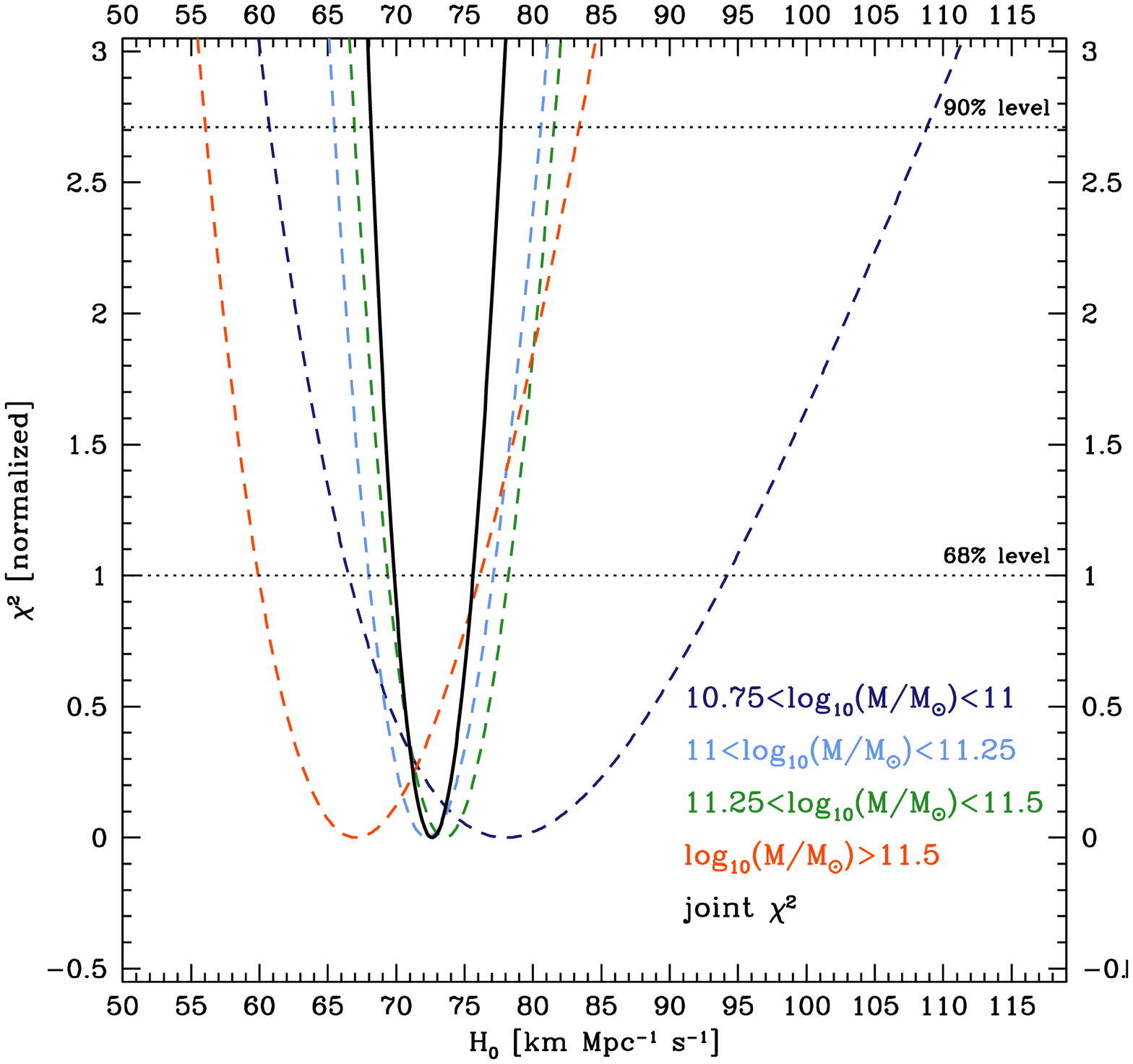}
\includegraphics[angle=0,width=0.49\textwidth]{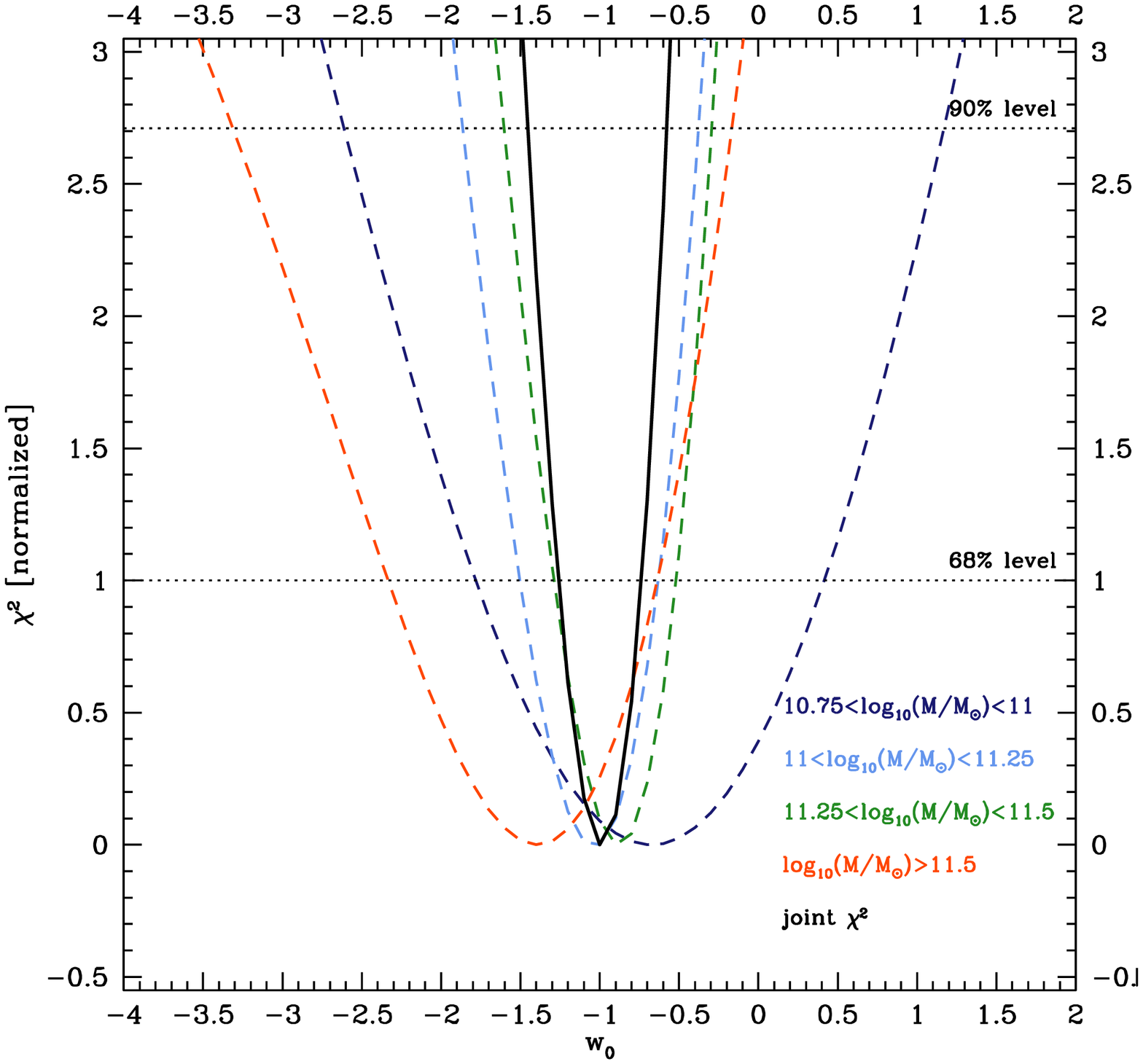}
}
\caption{$\chi^{2}$ of the $H_{0}$ (left panel) and of the $w$ (right panel) evaluation for ETGs in different mass subsamples using BC03 models.}
\label{fig:3}
\end{figure}
To estimate $H_{0}$, we fitted the $D4000_{n}$-z relations given in Fig. \ref{fig:1} with Eq. \ref{eq:3}, fixing as a prior only a flat $\Lambda CDM$ cosmology ($\Omega_{m}+\Omega_{DE}=1$, $w=-1$ and $w_{a}=0$), letting $\Omega_{DE}$ free to vary in the range provided by the latest WMAP 7-years analysis ($0.7<\Omega_{DE}<0.76$) and the slope A free to vary within the values associated to the metallicity range found in the data within $1\sigma$ error (see Table \ref{tab:1}), considering also the spread due to the different SFHs considered (see the second point of Sect. \ref{sec:syst}).\\
As explained in Sect. \ref{sec:stat}, the best-fit has been evaluated with a standard $\chi^{2}$ analysis, marginalizing the obtained likelihood over those parameters which we are not interested to constrain, that are $\Omega_{DE}$ and A. We end up with four likelihoods of the Hubble constant for the mass bins we created. The estimates obtained from the higher populated samples ($11<log(M/M_{\odot})<11.25$ and $11.25<log(M/M_{\odot})<11.5$) show a perfect agreement, with comparable errors; the analysis of the other two mass subsamples give slightly different results, but still within the $1\sigma$ errors. However, it is also important to remember that the number of galaxies of these subsamples is quite lower with respect to the bin $11<log(M/M_{\odot})<11.25$ (respectively 36\% and 72\% less), and therefore that the associated errors are larger (see \mbox{Tab. \ref{tab:3}}).

Since those likelihoods are independent from each other, to reduce the statistical error and to obtain a better estimate on $H_{0}$ we decided to join them, obtaining $H_{0}=72.6\pm2.9\;\mathrm{km\,Mpc^{-1}s^{-1}}$. In Fig. \ref{fig:3} are shown the $\chi^{2}$ for each mass bin and the joint $\chi^{2}$, and in Table \ref{tab:3} are reported the corresponding values of $H_{0}$ and their errors.
\begin{table}[b!]
\begin{center}
\begin{tabular}{ccccc}
\hline \hline
&$H_{0}$& 68\% errors & 90\% errors & \# galaxies\\
&$[\mathrm{km\,Mpc^{-1} s^{-1}}]$&$[\mathrm{km\,Mpc^{-1} s^{-1}}]$&$[\mathrm{km\,Mpc^{-1} s^{-1}}]$&\\
\hline
$10.75<log(M/M_{\odot})<11$&77.6&$+16.6$/$-11$&$+31.2$/$-16.8$&3452\\
$11<log(M/M_{\odot})<11.25$&72.2&$+4.9$/$-4.2$&$+8.3$/$-6.7$&5429\\
$11.25<log(M/M_{\odot})<11.5$&73.5&$+4.7$/$-4.1$&$+8$/$-6.5$&3591\\
$log(M/M_{\odot})>11.5$&66.9&$+9.2$/$-6.9$&$+16.4$/$-10.8$&1515\\
\hline \hline
joint analysis &72.6&$+3$/$-2.7$&$+5$/$-4.4$&13987\\
\end{tabular}
\caption{Hubble constant (in unit of $\mathrm{km\,Mpc^{-1} s^{-1}}$) and relative errors using BC03 models.}
\label{tab:3}
\end{center}
\end{table}
\begin{table}[b!]
\begin{center}
\begin{tabular}{ccccc}
\hline \hline
&$w$& 68\% errors & 90\% errors & \# galaxies\\
\hline
$10.75<log(M/M_{\odot})<11$&-0.7&$+1.1$/$-1.0$&$+1.8$/$-1.9$&3452\\
$11<log(M/M_{\odot})<11.25$&-1.0&$+0.3$/$-0.5$&$+0.6$/$-0.8$&5429\\
$11.25<log(M/M_{\odot})<11.5$&-0.9&$+0.3$/$-0.3$&$+0.6$/$-0.7$&3591\\
$log(M/M_{\odot})>11.5$&-1.4&$+0.7$/$-0.9$&$+1.2$/$-1.9$&1515\\
\hline \hline
joint analysis &-1&$+0.2$/$-0.2$&$+0.4$/$-0.4$&13987\\
\end{tabular}
\caption{Dark energy equation of state parameter $w$ and relative errors using BC03 models.}
\label{tab:5}
\end{center}
\end{table}

\subsection{The dark energy equation of state}
To estimate the value of $w$, assuming $w=\mathrm{constant}$, we fitted the $D4000_{n}$-z relations using as priors the $\Omega_{DE}$ value as found by the latest WMAP 7-years analysis considering a wCDM flat Universe (Larson et al. 2010, $\Omega_{m}+\Omega_{DE}=1$, $w_{a}=0$ and $0.645<\Omega_{DE}<0.835$) and the estimate of $H_{0}$ done by Riess et al. (2009) ($70.6\leq H_{0}\leq77.8$), the slope $A$ free to vary within the values associated to the metallicity range found in the data within $1\sigma$ error (see Table \ref{tab:1}), considering also the spread due to the different SFHs considered (see the second point of Sect. \ref{sec:syst}).\\
The best-fit has been evaluated with a standard $\chi^{2}$ analysis, marginalizing afterwards over $H_{0}$, $\Omega_{DE}$ and A. We end up with four likelihoods of $w$ for the mass bins we created. All the estimates obtained on $w$ are in agreement within $1\sigma$ error; only the value obtained in the highest mass bin shows a slight discrepancy, even if compatible with the other estimates within $1\sigma$. However, a similar discussion as the one introduced in the $H_{0}$ estimate has to be done for this discrepancy, considering that it is the bin with the highest errors and the lowest galaxy number. As done in the $H_{0}$ analysis, since those likelihood are independent one from the other we performed a joint analysis by multiplying them, to reduce the statistical error. In Fig. \ref{fig:4} are shown the $\chi^{2}$ for each mass bin and the joint $\chi^{2}$, and in Table \ref{tab:5} are reported the corresponding values of $w$ and their errors. From the joint analysis we obtained $w=-1\pm0.2$.

\subsection{Study of the systematics}
\label{sec:syst}
In our approach, we studied the effect of different possible systematics from which our analysis may suffer. In particular, the most critical ones are the dependence of our results on the estimate of the absolute age of a galaxy from the $D4000_{n}$, on the star formation histories considered, on the stellar population synthesis models assumed, on the metallicity evaluation and on the not perfect assumption of the passive evolution of our sample. In the following we address separately all these sources of uncertainty\\
\begin{itemize}
\item{\it Dependence on the absolute age evaluation.} Since our approach relies totally on the study of the differential age (or $D4000_{n}$) evolution to constrain cosmological parameters, our approach is insensitive to a possible bias in the evaluation of the absolute age of a galaxy.\\
    On the contrary, we strongly rely on the fact that our technique does not smear in some way the slope of the age-redshift evolution, i.e. that the differential evolution is preserved unmodified. This approach, by definition, keeps the differential evolution unchanged, since we use a simple linear conversion between the $D4000_{n}$ and the age of the galaxy, justifying in Sect. \ref{sec:model} the robustness of this assumption.
\end{itemize}
\begin{figure}[b!]
\mbox{
\includegraphics[angle=0,width=0.49\textwidth]{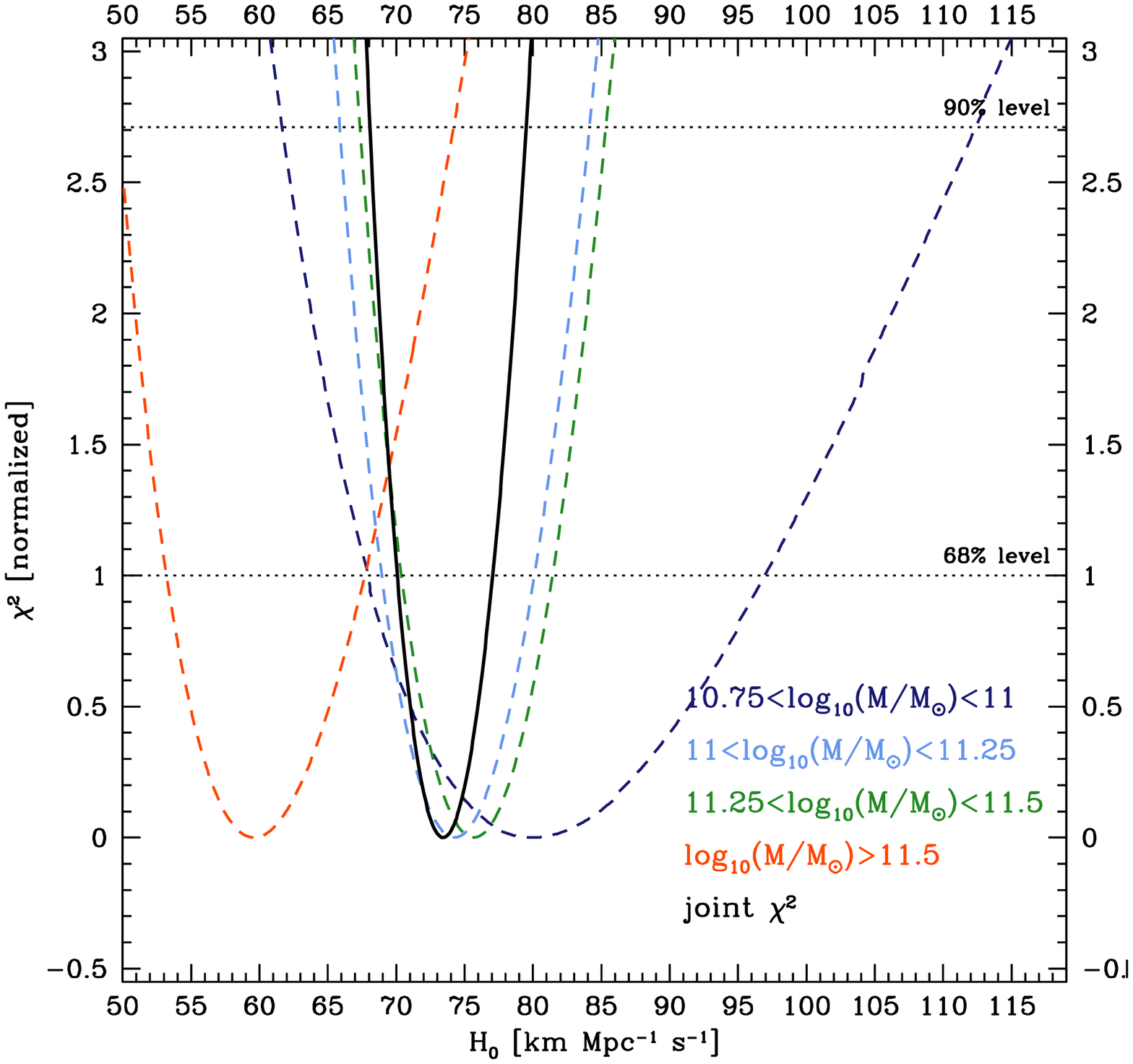}
\includegraphics[angle=0,width=0.49\textwidth]{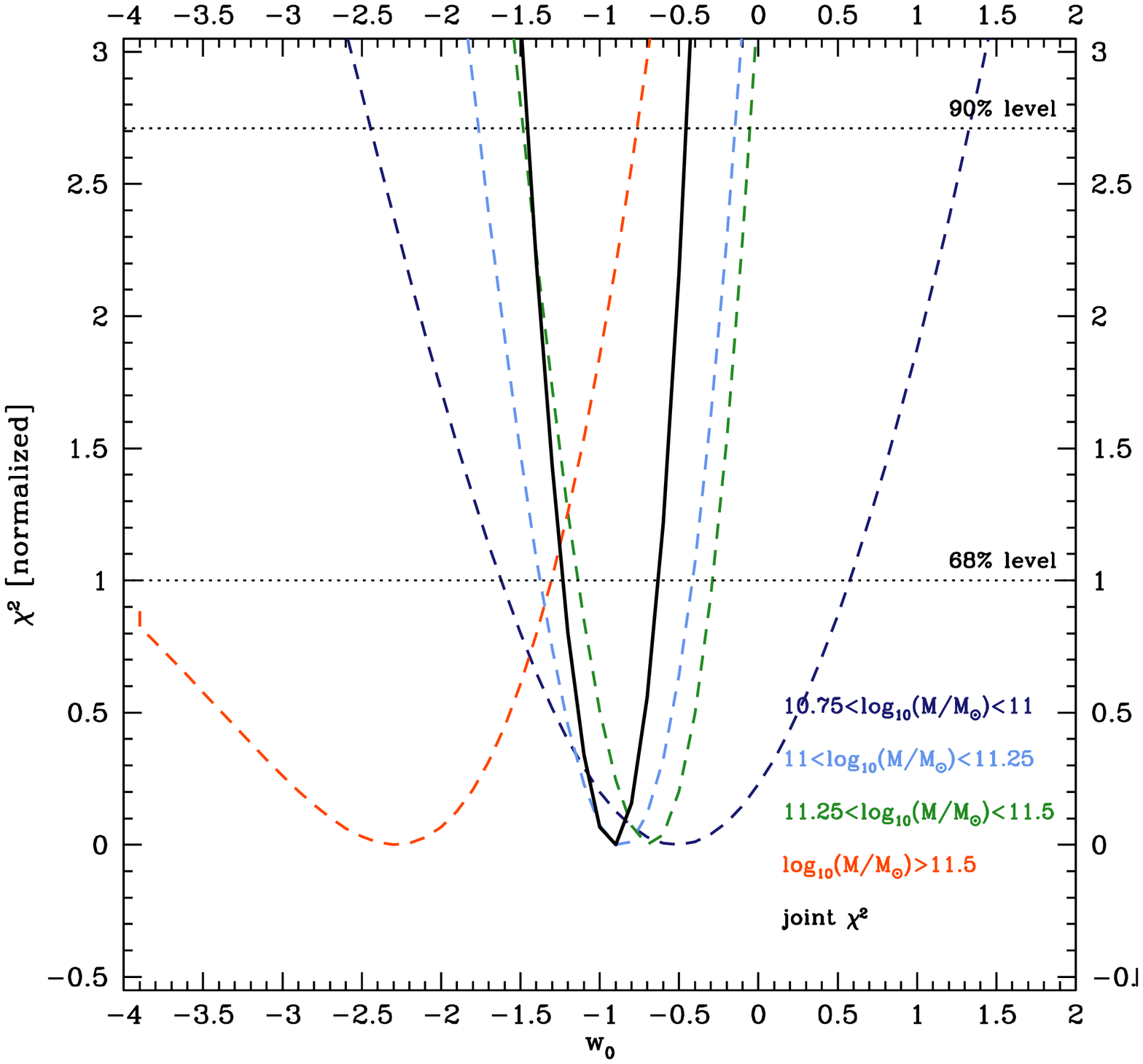}
}
\caption{$\chi^{2}$ of the $H_{0}$ (left panel) and of the $w$ (right panel) evaluation for ETGs in different mass subsamples using {\it Mastro} models.}
\label{fig:9}
\end{figure}
\begin{itemize}
\item{\it Dependence on the star formation histories considered.} In Sect. \ref{sec:model} we justify the choice of using delayed exponentially declining star formation histories with low $\tau$'s by analyzing the SED-fit tau distributions of our sample (see Fig. \ref{fig:6}). However, the study of models with different $\tau$'s introduces a spread in the conversion factor $A$ between the $D4000_{n}$ and the age of a galaxy, since we averaged the values found with the different SFHs at fixed metallicity. The percentage spread in the slope, averaged in our metallicity range, is $3.1\%$ and $2.8\%$ for BC03 models respectively in the {\it "high D4000"} regime and $13.4\%$ in the {\it "low D4000"} regime), and $3.9\%$ and $12.4\%$ for {\it Mastro} models, and the effect is even smaller (almost negligible) in the estimated cosmological parameters. Even if this spread is small, we take in consideration this additional error in our analysis.
\end{itemize}
\begin{table}[t!]
\begin{center}
\begin{tabular}{ccccc}
\hline \hline
&$H_{0}$& 68\% errors & 90\% errors & \# galaxies\\
&$[\mathrm{km\,Mpc^{-1} s^{-1}}]$&$[\mathrm{km\,Mpc^{-1} s^{-1}}]$&$[\mathrm{km\,Mpc^{-1} s^{-1}}]$&\\
\hline
$10.75<log(M/M_{\odot})<11$&79.5&$+17.4$/$-11.5$&$+32.8$/$-17.8$&3452\\
$11<log(M/M_{\odot})<11.25$&74.1&$+6$/$-5.1$&$+10$/$-8.2$&5429\\
$11.25<log(M/M_{\odot})<11.5$&75.6&$+5.8$/$-5.2$&$+9.7$/$-8.2$&3591\\
$log(M/M_{\odot})>11.5$&59.4&$+8.3$/$-6.1$&$+14.7$/$-9.3$&1515\\
\hline \hline
joint analysis &73.4&$+3.6$/$-3.3$&$+6.1$/$-5.3$&13987\\
\end{tabular}
\caption{Hubble constant (in unit of $\mathrm{km\,Mpc^{-1} s^{-1}}$) and relative errors using {\it Mastro} models.}
\label{tab:7}
\end{center}
\end{table}
\begin{table}[t!]
\begin{center}
\begin{tabular}{ccccc}
\hline \hline
&$w$& 68\% errors & 90\% errors & \# galaxies\\
\hline
$10.75<log(M/M_{\odot})<11$&-0.5&$+1.0$/$-1.1$&$+1.8$/$-1.9$&3452\\
$11<log(M/M_{\odot})<11.25$&-0.9&$+0.4$/$-0.4$&$+0.7$/$-0.8$&5429\\
$11.25<log(M/M_{\odot})<11.5$&-0.7&$+0.4$/$-0.4$&$+0.6$/$-0.7$&3591\\
$log(M/M_{\odot})>11.5$&-2.3&$+0.9$/$-1.6$&$+1.5$/$-1.6$&1515\\
\hline \hline
joint analysis &-0.9&$+0.2$/$-0.3$&$+0.4$/$-0.5$&13987\\
\end{tabular}
\caption{Dark energy equation of state parameter $w$ and relative errors using {\it Mastro} models.}
\label{tab:8}
\end{center}
\end{table}
\begin{itemize}
\item{\it Dependence on the choice of the stellar population synthesis model.} In Sect. \ref{sec:model} we perform the analysis of the slopes of the $D4000_{n}$-age relation both using BC03 and {\it Mastro} models. As pointed out before, these two models are completely different, being the {\it Mastro} models based on the latest MILES models and the BC03 models on the STELIB ones. In \mbox{Tab. \ref{tab:2}} we show the values of the slopes in the two $D4000_{n}$ regimes for the two models. To quantify the difference between these two model, since we are interested in our particular metallicity range shown in \mbox{Tab. \ref{tab:1}}, we interpolate the values of the slope $<A>$-metallicity relation in \mbox{Tab. \ref{tab:2}} for both models with a quadratic function. We find that the mean percentage difference between the slope obtained from the two models in our metallicity range is $2.8\%$ in the {\it "low D4000"} regime and $13.4\%$ in the {\it "high D4000"} regime.\\
We also evaluated the effect of this percentage difference between the models by performing the same analysis described in Sect. \ref{sec:H0} with {\it Mastro} models. The values of $H_{0}$ and $w$ obtained in the different mass bins are shown in \mbox{Tab. \ref{tab:7}} and \mbox{Tab. \ref{tab:8}}, and the corresponding $\chi^{2}$ are shown in Fig. \ref{fig:9}. After marginalizing and doing the joint analysis combining the informations of all the mass bins, we find using {\it Mastro} models $H_{0}=73.4\pm3.5\;\mathrm{km Mpc^{-1} s^{-1}}$ and $w=-0.9\pm0.3$. We find therefore no significant difference between BC03 and {\it Mastro} models, obtaining values for both $H_{0}$ and $w$ compatible within the $1\sigma$ errors.
\end{itemize}
\begin{itemize}
\item{\it Dependence on metallicity and on star formation.} An important test of this analysis is to verify the robustness of our results against the choice of the value of the metallicity, since it is the primary parameter to obtain the correct slope of the $D4000_{n}$-age relation. To check that our analysis is not biased in that sense, we studied the mass bin with the highest number of galaxies ($11<log(M/M_{\odot})<11.25$), dividing it into two subsample sampling different metallicity ranges; we decide to divide the sample below its median metallicity ({\it "low metallicity"} sample) and above its median metallicity ({\it "high metallicity"} sample), to keep the number size of the resulting subsamples similar. In this way we end up with two subsample with significantly different values of metallicity ($Z/Z_{\odot}=0.92$ and $Z/Z_{\odot}=1.38$, respectively).
\begin{figure}[t!]
\begin{centering}
\mbox{\includegraphics[angle=-90,width=0.75\textwidth]{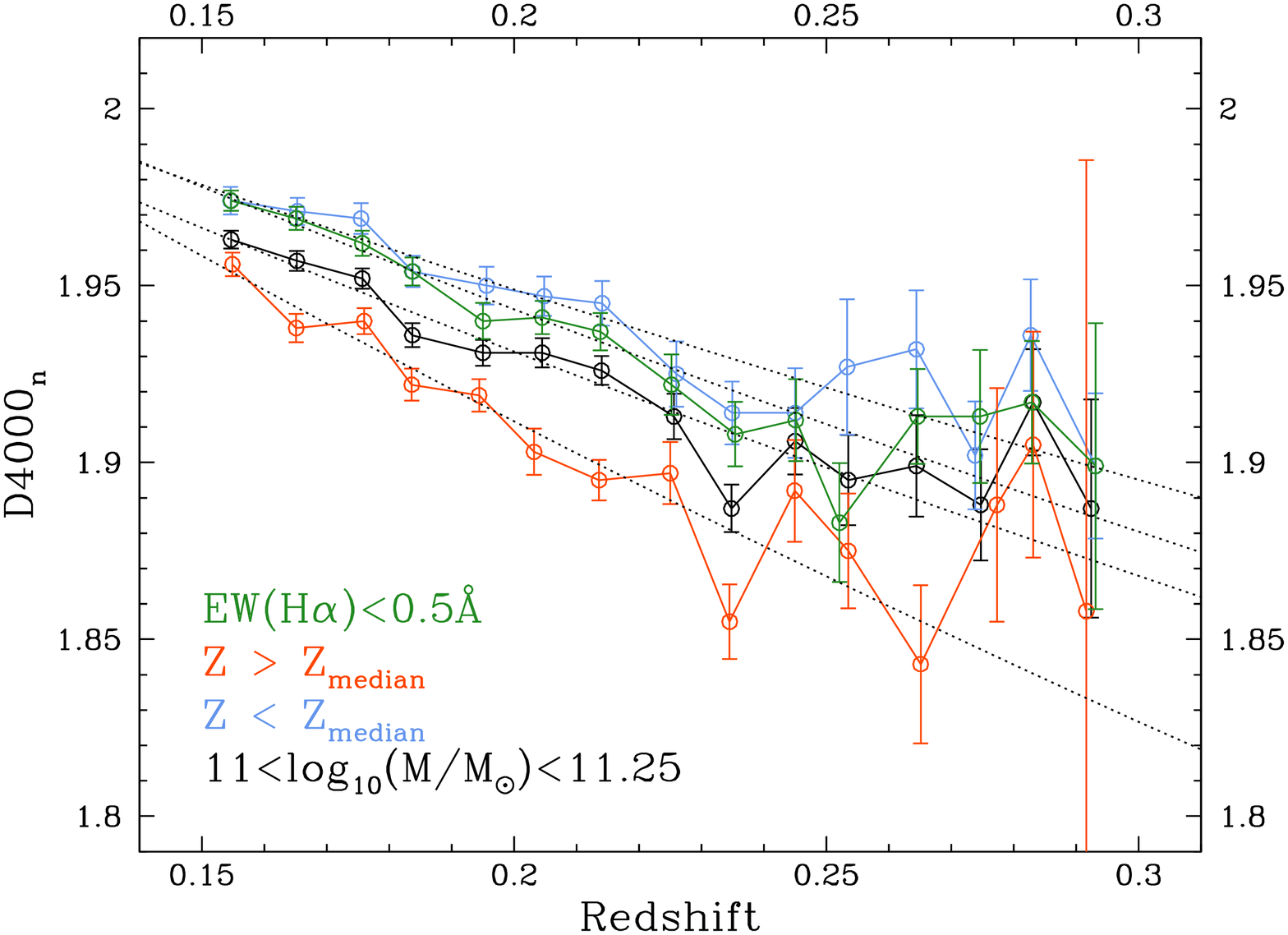}}
\mbox{
\includegraphics[angle=0,width=0.45\textwidth]{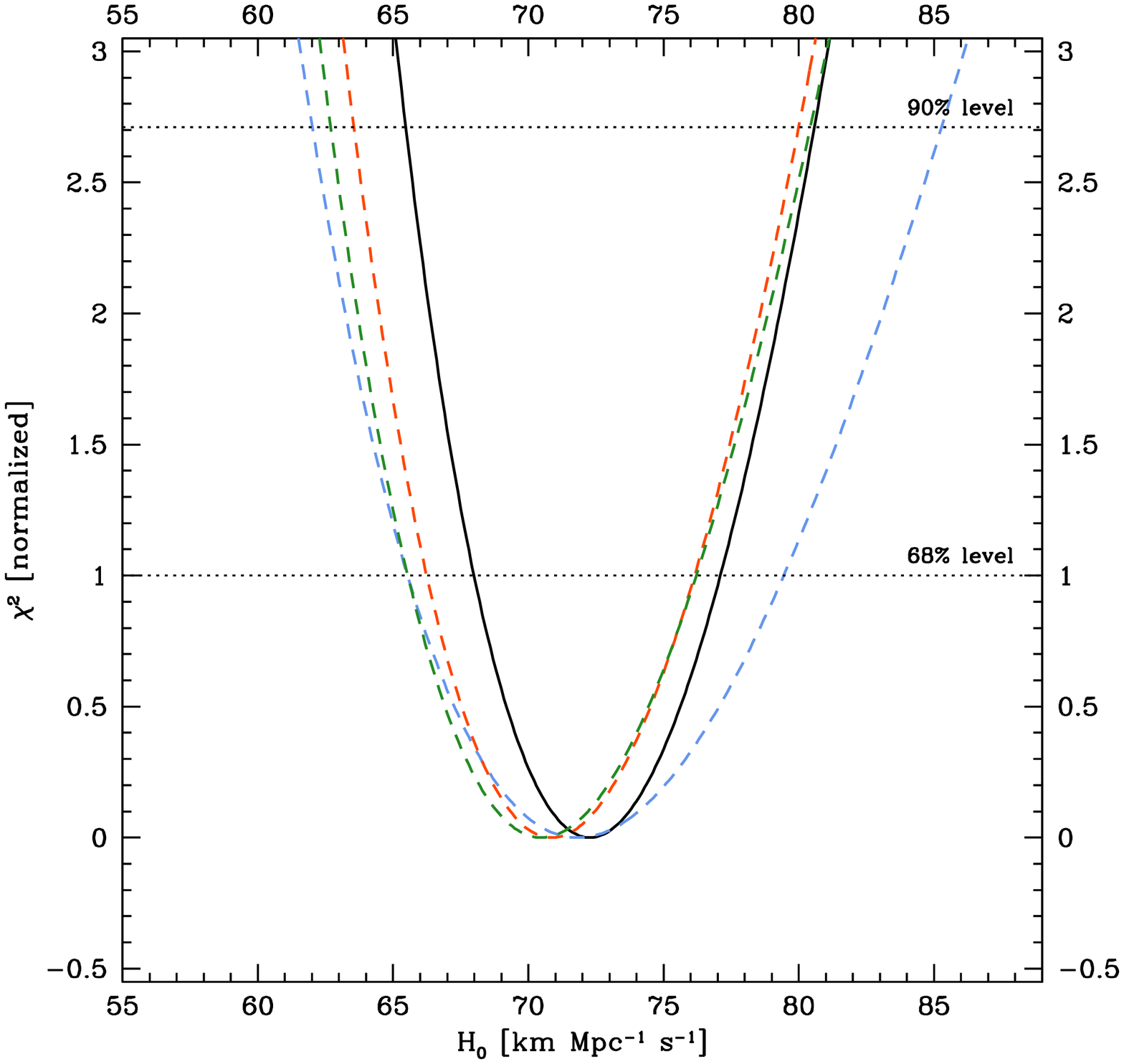}
\includegraphics[angle=0,width=0.45\textwidth]{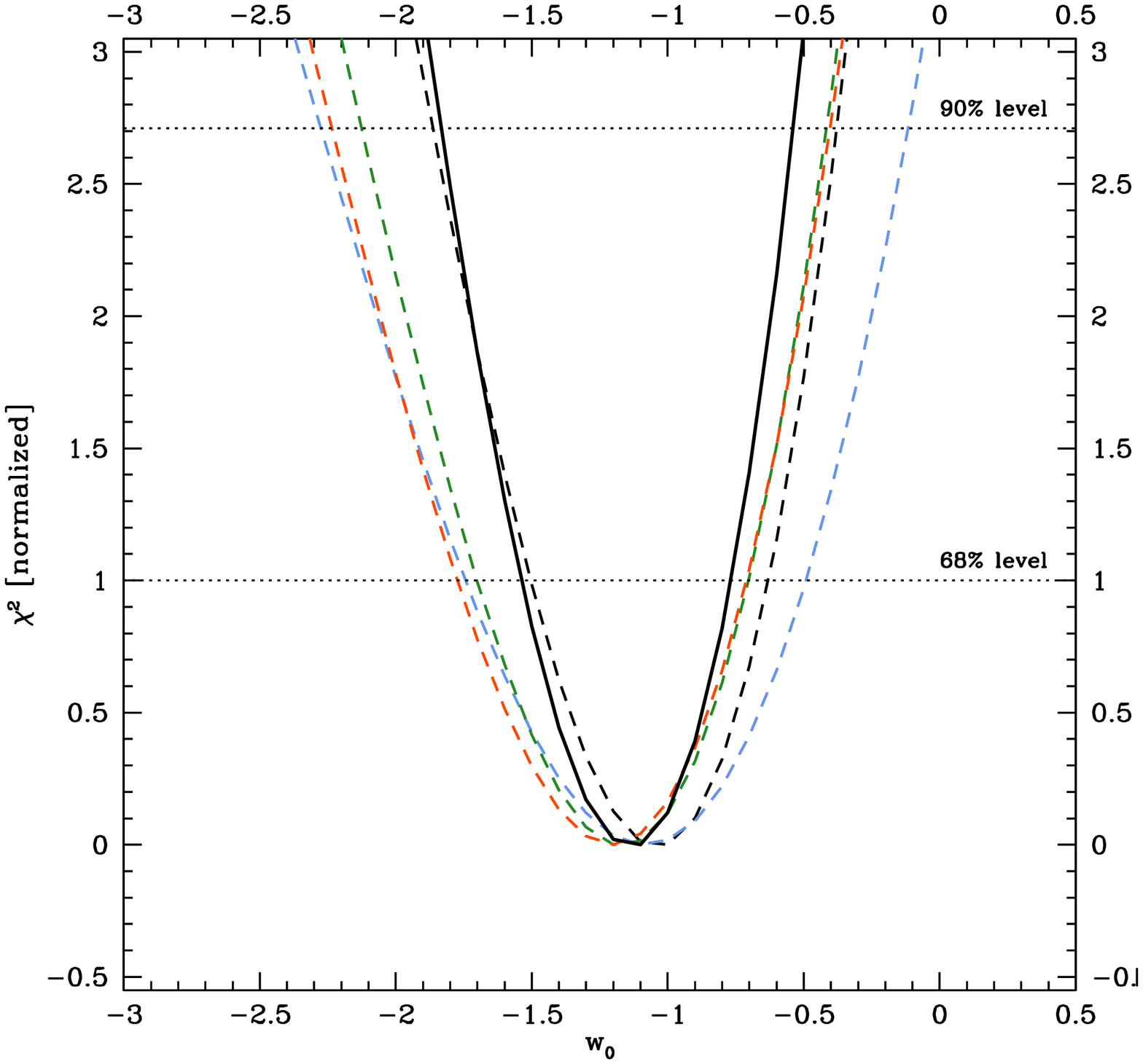}
}
\caption{In the left panel is shown the D4000-redshift relation for ETGs with \mbox{$11<log(M/M_{\odot})<11.25$} (in black), for the same mass bin sample divided below the median metallicity (in blue) and above the median metallicity (in red), and for the same mass bin selected with $EW_{0}(H\alpha)<0.5$ {\AA} (in green). In the right panel there are the corresponding $\chi^{2}$.}
\label{fig:4}
\end{centering}
\end{figure}
\begin{table}[t!]
\begin{center}
\begin{tabular}{ccccc}
\hline
&$H_{0}$& 68\% errors & 90\% errors & \# galaxies\\
& $\mathrm{km\,Mpc^{-1}s^{-1}}$ & $\mathrm{km\,Mpc^{-1}s^{-1}}$ & $\mathrm{km\,Mpc^{-1}s^{-1}}$\\
\hline
$11<log(M/M_{\odot})<11.25$&72.2&+4.9/-4.2&+8.3/-6.7&5429\\
\hline
{\it "low Z"}&71.7&+7.7/-6.1&+13.5/-9.6&2717\\
{\it "high Z"}&70.8&+5.3/-4.5&+9.1/-7.2&2712\\
$EW_{0}(H\alpha)<0.5$\AA& 70.4 &+5.8/-4.8&+10/-7.7&3245\\
\hline
\end{tabular}
\caption{Hubble constant (in unit of $\mathrm{km\,Mpc^{-1} s^{-1}}$) and relative errors for \mbox{$11<log(M/M_{\odot})<11.25$} sample, for the same mass bin sample divided below the median metallicity, above the median metallicity and selected with $EW_{0}(H\alpha)<0.5$\AA, using BC03 models.}
\label{tab:4}
\end{center}
\end{table}

\begin{table}[t!]
\begin{center}
\begin{tabular}{ccccc}
\hline
&$w$& 68\% errors & 90\% errors & \# galaxies\\
\hline
$11<log(M/M_{\odot})<11.25$&-1&+0.3/-0.5&+0.6/-0.8&5429\\
\hline
{\it "low Z"}&-1.1&+0.5/-0.6&+0.9/-1.1&2717\\
{\it "high Z"}&-1.2&+0.4/-0.5&+0.7/-0.9&2712\\
$EW_{0}(H\alpha)<0.5$\AA& -1.2 &+0.4/-0.5&+0.7/-1&3245\\
\hline
\end{tabular}
\caption{Dark energy equation of state $w$ and relative errors for \mbox{$11<log(M/M_{\odot})<11.25$} sample, for the same mass bin sample divided below the median metallicity, above the median metallicity and selected with $EW_{0}(H\alpha)<0.5$\AA, using BC03 models.}
\label{tab:10}
\end{center}
\end{table}

    In Fig. \ref{fig:4} (upper panel) it is shown the $D4000_{n}$-z relation of the parent sample (in black) and of the {\it "low metallicity"} and {\it "high metallicity"} subsamples (respectively in blue and red). We evaluated the median metallicity of the two subsamples, to obtain for each one the correct A parameter. The corresponding $D4000_{n}$-redshift relations result rather different between the {\it "low metallicity"} and the {\it "high metallicity"} sample; however given the different metallicity and therefore conversion parameter A, the fit to $H_{0}$ and $w$ gives results in perfect agreement, as it is possible to see from the $\chi^{2}$ in the lower panels of Fig. \ref{fig:4} and from the values reported in Table \ref{tab:4} and Table \ref{tab:10}.\\
    Furthermore, from the analysis of the emission lines of our ETGs, we found that, in spite of the rather strict selection criterium applied, we still have a tail in the $EW_{0}(H\alpha)$ and $EW_{0}([OII]\lambda3727)$. On the one hand, as pointed out in Sect. \ref{sec:massmet}, these are quite low levels of equivalent widths, and the distributions of the \textsf{eClass} parameter and the specific star formation rate obtained from the SED fitting for the galaxies with a significant detection are characteristic of a passive population. On the other hand, this may be an indication that our sample contains a certain number of galaxies with undergoing star formation or AGN activity. Our approach relies strongly on the assumption that we are selecting a sample of passively evolving galaxies, that traces uniformly the age evolution of the Universe as a function of redshift. To test if there is a such kind of bias in our results, we decided to apply to the same mass subsample chosen before ($11<log(M/M_{\odot})<11.25$) an even stricter selection, considering only those galaxies for which $EW_{0}(H\alpha)<0.5$ {\AA} (that is the median of the $EW_{0}(H\alpha)$ distribution). We evaluated the metallicity of this sample, and redid the analysis of the Hubble constant and dark energy equation of state parameter, and the results are shown in Fig. \ref{fig:4} (green lines), and \mbox{Tab. \ref{tab:4}} and \mbox{Tab. \ref{tab:10}}. We find that the value obtained from this analysis well agrees with the value obtained not cutting the sample in $EW_{0}(H\alpha)$. So we conclude that our results are robust also against the presence of galaxies with small values of $EW_{0}(H\alpha)$.
\end{itemize}

\begin{table}[t!]
\begin{center}
\small
\begin{tabular}{cccc}
\hline
{\bf Source of uncertainty} & {\bf Effect} & {\bf Impact on} & {\bf Impact on}\\
{\bf uncertainty} & & {\bf $H_{0}$} & {\bf w}\\
\hline
Absolute age evaluation & possible underestimate of the & none & none\\
 & absolute ages using $D4000_{n}$ &  &\\
\hline
SFHs assumption & spread in the conversion factor A & $<0.01\%$ & $<0.1\%$\\
\hline
Stellar population & different $D4000_{n}$-age relation & $1.1\%$ & $10\%$\\
synthesis model used &  within models & & \\
\hline
Metallicity estimate & if incorrect, may bias the $A$ parameter & $1.4\%$ & 15\%\\
\hline
non-perfect approximation & possible bias in the $D4000_{n}$-z evolution & $2.5\%$ & 20\%\\
of passive evolution & & & \\
\hline
{\bf Total systematic error} & & {\bf $3.1\%$} & {\bf $27\%$}\\
\end{tabular}
\normalsize
\caption{Summary of the sources of uncertainty and their impact (in percentage) on the cosmological parameters.}
\label{tab:9}
\end{center}
\end{table}

\section{Total systematic uncertainty and results}
In Sect. \ref{sec:syst} we quantified the impact of systematics on the estimate of both the Hubble parameter $H_{0}$ and the dark energy equation of state parameter $w$ (assuming $w=\mathrm{constant}$). All these source of uncertainties, along with their effect, are summarized in \mbox{Tab. \ref{tab:9}}; to evaluate the effect of systematics, most of the tests have been performed using the mass sample with the highest number of galaxies ($11<log(M/M_{\odot})<11.25$). The total estimated effect has been evaluated by summing in quadrature the errors, and it is also reported in \mbox{Tab. \ref{tab:9}}.

In conclusion, we find that our estimated value for $H_{0}$, including both statistic and systematic errors, is \mbox{$H_{0}=72.6\pm2.9(stat)\pm2.3(syst)\:\mathrm{km\,Mpc^{-1}s^{-1}}$}; our estimated value for $w$, including both statistic and systematic errors, is \mbox{$w=-1\pm0.2(stat)\pm0.3(syst)$}.


\section{Summary and conclusions}
\label{sec:conclusions}
We have developed a new technique to obtain the expansion rate of the Universe using the ``cosmic chronometer'' technique with the aim of minimizing the dependence on systematics. To this extent, we have shown that the $D4000_{n}$ feature at fixed metallicity correlates linearly with age, for the range of ages of interest. We have studied this feature using theoretical synthetic stellar population models and shown that it is robust to choice of metallicity, star formation history and different stellar population models. We also evaluated, within this linear approximation, the theoretical model $D4000_{n}(z)$ as a function of the various cosmological parameters.

We obtained a sample of ETGs from the SDSS survey, using both photometric and spectroscopic information to select the most passive and massive galaxies ($log(M/M_{\odot})>10.75$). We divide our sample in small mass bins (0.25 dex), in order to avoid possible bias from mass downsizing and to have homogeneous redshift of formation and metallicity across the entire redshift range. The $D4000_{n}-z$ relations show, in each subsample, a clear redshift evolution, with galaxies at low redshift having always a higher $D4000_n$, and hence higher ages. Moreover the four subsamples give an evident proof of mass-downsizing, since at each redshift galaxies with higher masses have always an higher $D4000_{n}$ than galaxies at lower masses.

We evaluate the Hubble constant $H_{0}$ and the dark energy equation of state parameter $w$ by fitting the $D4000_{n}-z$ relations with the theoretical model defined before. We find from the joint analysis that \mbox{$H_0 = 72.6 \pm 2.9(stat)\pm2.3(syst)\:\mathrm{km\,Mpc^{-1}s^{-1}}$} and that \mbox{$w = -1 \pm 0.2(stat)\pm0.3(syst)$}, assuming a constant $w$. The value we obtained for $w$ is in agreement with the value found from the WMAP 7-years analysis and with the one obtained from Amanullah et al. (2010); the value of $H_0$ is within $1\sigma$ error with the one obtained by Riess et al. (2009), and we reach a comparable level of precision. We have also shown that the universe is accelerating, even if our sample only reaches up to $z \sim 0.3$.

We demonstrate that our results are robust to the choice of galaxy mass, since the results in the single mass subsample are all within the $1\sigma$ errors. Moreover, we study the effect of the dependence of our results on different systematics, i.e. the estimate of the absolute age of a galaxy from the $D4000_{n}$, the star formation histories considered, the stellar population synthesis models assumed, the metallicity evaluation and the not perfect assumption of the passive evolution of our sample. We find a small dependence on the SFHs studied, that we took into account in the cosmological parameters evaluation. By studying our sample with different synthetic stellar populations models (BC03 and {\it Mastro}), in different metallicity regimes and for different selection of $EW_{0}(H\alpha)$, we show that our analysis is also robust to the metallicity evaluation and to the non-perfect approximation of passive evolution of our sample, and that there is no significant difference when using different models for studying the $D4000_{n}$-age relation.

In forthcoming papers we will apply our method to higher redshift samples, to determine the expansion history of the Universe, i.e. H(z), up to $z \sim 1$. This method is promising also in view of the future massive spectroscopic surveys expected in the next decades from the ground (e.g. SDSS-III BOSS, BigBOSS) and from space (e.g. ESA Euclid, NASA WFIRST).


\section*{Acknowledgments}
The authors would like to thank the anonymous referee for the constructive comments, Anna Gallazzi, Mariangela Bernardi, Marcella Carollo, Simon Lilly, Claudia Maraston, Daniel Thomas and Giovanni Zamorani for the useful discussions, and Adam Riess for the helpful comments. We would also like to thank Claudia Maraston for providing new stellar population synthesis models ({\it Mastro}) to study the D4000-age relation and Jarle Brinchmann for providing the starting sample of SDSS galaxies. AC, MM and LP acknowledge the support from ASI through the contract ``Euclid-NIS'' (I/039/10/0), and from MIUR with PRIN-2008 {\it ``Energia oscura e cosmologia con grandi survey di galassie''}.


\section*{References}
\bibliographystyle{JHEP}

\providecommand{\href}[2]{#2}\begingroup\raggedright

\end{document}